\input harvmac
\input epsf
\input poincare.defs

\overfullrule=0mm \hfuzz 10pt

\catcode`\@=11 
\def\eqna#1{\DefWarn#1\wrlabeL{#1$\{\}$}%
\xdef #1##1{(\noexpand\relax\noexpand\checkm@de%
{\s@csym\the\meqno\noexpand\f@rst{##1}}{\hbox{$\secsym\the\meqno##1$}})}
\global\advance\meqno by1}
\catcode`\@=12 

\font\ninerm=cmr9
\font\ninei=cmmi9
\font\nineb=cmbx9
\font\nineib=cmmib9
\font\sixi=cmmi6
\def\ninemath{
\textfont0=\ninerm
\textfont1=\ninei \scriptfont1=\sixi \scriptscriptfont1=\scriptfont1}

\def\fig#1#2#3{
\nobreak
\par\begingroup\parindent=0pt\leftskip=1cm\rightskip=1cm\parindent=0pt
\baselineskip=12pt \ninerm \ninemath
\midinsert
\centerline{\hbox{#3}}
\vskip 6pt
{\nineb Fig.~#1:}~#2\par
\vskip 6pt
\endinsert
\endgroup\par
\goodbreak
}

\def\({\left(}  \def\){\right)}
\def\[{\left[}  \def\]{\right]}
\def\h{{1\over2}}
\def\th{\textstyle{1\over2}}
\def\ra{\rightarrow}
\def\bra{\langle}       \def\ket{\rangle}
\def\<{\langle}
\def\>{\rangle}

\def\b#1{\bar{#1}}
\def\frc#1#2{{#1 \over #2}}
\def\dist{{\rm d}}
\def\lt#1{\left#1}
\def\rt#1{\right#1}
\def\d{\partial}
\def\Or{{\cal O}}
\def\mR{r}

\def\rx{{\rm x}}
\def\ry{{\rm y}}

\def\cov{{\cal D}}
\def\cov{{\cal D}}
\def\P{{\bf P}}
\def\Pbar{{\bf\bar P}}
\def\R{{\bf R}}
\def\X{{\bf X}}
\def\Y{{\bf Y}}
\def\iline{{\cal C}}

\def\psiiso{\psi^{(\rm iso)}}
\def\bpsiiso{\bar\psi^{(\rm iso)}}
\def\vac{{\rm vac}}

\def\Ione{\hbox{$1\hskip -1.2pt\vrule depth 0pt height 1.53ex width 0.7pt
                  \vrule depth 0pt height 0.3pt width 0.12em$}}
\def\strucid{
C_{1\hskip -0.9pt\vrule depth 0pt height 1.05ex width 0.48pt
\vrule depth 0pt height 0.3pt width 0.08em}}

\font\cmss=cmss10 \font\cmsss=cmss10 at 7pt
\def\IZ{\relax\ifmmode\mathchoice
{\hbox{\cmss Z\kern-.4em Z}}{\hbox{\cmss Z\kern-.4em Z}}
{\lower.9pt\hbox{\cmsss Z\kern-.4em Z}} {\lower1.2pt\hbox{\cmsss
Z\kern-.4em Z}}\else{\cmss Z\kern-.4em Z}\fi}
\def\IR{{\rm I}\!{\rm R}}

\def\pre#1{{\tt \hbox{#1}}}

\lref\Cardy{J.~Cardy, {\sl Conformal invariance and surface
critical behaviour}, Nucl.~Phys. B 240 (1984) 514--532.}

\lref\BreitFreedman{P. Breitenlohner, D. Z. Freedman, {\sl
Positive energy in anti-de Sitter backgrounds and gauged extended
supergravity}, Phys. Lett. B 115 (1982) 197--201.}

\lref\Tracy{J.~Palmer, M.~Beatty, C.~A.~Tracy, {\sl Tau functions
for the Dirac operator on the Poincar\'e disk},
Commun.~Math.~Phys. 165 (1994) 97--174, \pre{hep-th/9309017}.}

\lref\DoyonI{B.~Doyon, {\sl Two-point correlation functions of
scaling fields in the Dirac theory on the Poincar\'e disk}, Nucl.
Phys. B 675 (2003) 607--630, \pre{hep-th/0304190}.}

\lref\DoyonII{B.~Doyon, {\sl Form factors of Ising spin and
disorder fields on the Poincar\'e disk}, J. Phys. A 37 (2003)
359-370, special edition, proceedings of the workshop {\sl Recent
Advances in the Theory of Integrable Systems} (Annecy, 2003).}

\lref\FZamoI{P.~Fonseca, A.~Zamolodchikov, {\sl Ising Field
Theory in a Magnetic Field: Analytic Properties of the Free Energy},
J.~Stat.~Phys. 110 (2003) 527--590, \pre{hep-th/0112167}.}

\lref\FonsecaZamoII{P.~Fonseca, A.~Zamolodchikov, {\sl Ward identities and
integrable differential equations in the Ising field theory},
\pre{hep-th/0309228}.}

\lref\Drouffe{C.~Itzykson, J.--M.~Drouffe, Statistical field
theory, Cambridge University Press, 1989.}

\lref\Luther{A.~Luther, I.~Peschel, {\sl Calculation of critical
exponents in two dimensions from quantum field theory in one
dimension}, Phys.~Rev.~B 12 (1975) 3908--3917.}

\lref\Rietman{R. Rietman, B. Nienhuis, J. Oitmaa, {\sl The Ising
model on hyperlattices}, J.~Phys. A~25 (1992) 6577--6592.}

\lref\Kadanoff{L. P. Kadanoff, H. Ceva, {\sl Determination of an
operator algebra for the two-dimensional Ising model}, Phys. Rev.
B 3 (1971) 3918--3938.}

\lref\AlZamo{Al. B. Zamolodchikov, {\sl Scaling Lee-Yang model on
a sphere I. Partition function}, JHEP 0207:029 (2002),
\pre{hep-th/0109078}.}

\lref\CardyBook{J.~Cardy, Scaling and renormalization in
statistical physics, Cambridge University Press, 1996.}

\lref\Callan{C. G. Callan, F. Wilczek, {\sl Infrared behaviour at
negative curvature}, Nucl.~Phys, B~340 (1990) 366--386.}

\lref\Leclair{A.~LeClair, {\sl Spectrum generating affine Lie algebras
in massive field theory}, Nucl.~Phys. B 415 (1994) 734--780,
\pre{hep-th/9305110}.}

\lref\Wu{T. T. Wu, B. M. McCoy, C. A. Tracy, E. Barouch, {\sl
Spin-spin correlation functions for the two-dimensional Ising model:
exact theory in the scaling region}, Phys.~Rev. B~13 (1976) 316--374.}

\Title
{\vbox{\baselineskip12pt
\hbox{RUNHETC-2003-37}
\hbox{SPhT-T04/031}
}}
{\vbox{\centerline{Ising Field Theory on a Pseudosphere}}}
\smallskip
\centerline{Benjamin Doyon$^1$ and Pedro Fonseca$^2$}

\medskip
\bigskip

\centerline{\sl $^1$ Department of Physics and Astronomy, Rutgers University}
\centerline{\sl Piscataway, NJ 08855-0849, USA}
\centerline{\tt doyon@physics.rutgers.edu}

\medskip
\bigskip

\centerline{\sl $^2$ Service de Physique Th{\'e}orique, CEA Saclay}
\centerline{\sl F-91191 Gif-sur-Yvette CEDEX, France}
\centerline{\tt pfonseca@spht.saclay.cea.fr}

\vskip 0.6in

\noindent We show how the symmetries of the Ising field theory on a
pseudosphere can be exploited to derive the form factors of the spin
fields as well as the non-linear differential equations satisfied by the
corresponding two-point correlation functions. The latter are studied in
detail and, in particular, we present a solution to the so-called
connection problem relating two of the singular points of the associated
Painlev\'e VI equation. A brief discussion of the thermodynamic properties
is also presented.

\Date{4/2004}

\newsec{Introduction}

The study of the behaviour of statistical systems on a curved
space is a problem with much room for development. A natural
question concerns the modification of the critical properties of a
statistical system due to a nonzero curvature, which introduces an
additional scale. This question is of interest both for its
relations to quantum gravity, and for understanding the
thermodynamics of classical systems on a curved space, a point of
view which we adopt in this paper. It is natural to expect that
the study of quantum field theories on a curved
Euclidean-signature space can lead to a better understanding of
this problem. In this paper we are interested in studying the
Ising field theory at zero magnetic field on a two-dimensional
curved space of constant negative curvature (the pseudosphere).
The interest in this space is in part technical: since a space of
constant curvature is maximally symmetric, known techniques
developed for the case of flat space can be extended to the Ising
field theory on such a space. The pseudosphere also has unusual
characteristics, for instance it has an infinite (two-dimensional)
volume while providing an infrared regularization, and it can be
expected to have nontrivial effects on the thermodynamics. The
study of these effects should in fact throw light on the main
properties of the thermodynamics on negatively curved spaces.

We will be mainly interested in studying the two-point
correlation functions of the order and disorder fields $\sigma(x)$ and
$\mu(x)$ in the Ising field theory on the pseudosphere. Although
this theory possesses stable regimes where symmetries associated
to some of the isometries of the pseudosphere are broken (as we
briefly explain in Section 2), we will only consider regimes with
unbroken symmetries. Using the parametrization of the pseudosphere on the
unit disk (the Poincar\'e disk), the theory can be described in terms of
the boundary Ising conformal field theory \Cardy\ deformed by the
energy field, with action
\eqn\action{
{\cal A} = {\cal A}_{\rm BI}
- {m\over2\pi}\int_{\rm disk} d^2x~ e^{\phi(x)} \,\varepsilon(x)~.
}
Here, ${\cal A}_{\rm BI}$ stands for the action of the Ising
conformal field theory on the unit disk and $d^2x\,
e^{\phi(x)}$ is the volume element in the metric of the
Poincar\'e disk (written explicitly in \metric). The energy
field $\varepsilon(x)$ is normalized by
\eqn\energynorm{
\dist(x,x')^2~\bra\varepsilon(x)\varepsilon(x')\ket\ra1
\indent\hbox{as}\indent
\dist(x,x')\ra0~,
}
where $\dist(x,x')$ is the geodesic distance between the points
$x$ and $x'$. Because we have in mind an Ising statistical system
on a curved space (by opposition to a system on a space with
boundaries), our definitions of the spin and disorder fields
differ from the usual ones defined in the Ising conformal field
theory on the disk, in the sense that the fields $\varepsilon(x)$,
$\sigma(x)$ and $\mu(x)$ are chosen scalar under the isometry
group of the pseudosphere.

The boundary Ising conformal field theory ${\cal A}_{\rm BI}$
admits ``free'' and ``fixed'' boundary conditions \Cardy, whereby
the order field has, respectively, zero and nonzero vacuum
expectation value\foot{With the ``fixed'' boundary condition, the
order field can be fixed to a positive or a negative value at the
boundary; we will choose to fix it to a positive value throughout
the paper.} (and vice-versa for the disorder field). Likewise,
introducing the parameter~$R$ related to the Gaussian curvature $\hat R$
by
\eqn\curvature{
\hat R = -{1\over R^2}~,
}
it is possible to see that when the energy perturbation is turned
on ($m\neq0$), the resulting theory \action\ on the pseudosphere
possesses in the region $-\frc12<mR<\frc12$ stable asymptotic
conditions corresponding to ``free'' and ``fixed'' conditions: the
order field has, respectively, zero and nonzero vacuum expectation
value, and vice-versa for the disorder field. In the domain
$mR>\frc12$ only the ``fixed'' asymptotic condition is stable,
whereas in the domain $mR<-\frc12$ only the ``free'' asymptotic
condition is stable (this description is in close connection with
some results of, for instance, Ref.~\BreitFreedman). Situations
with ``free'' and ``fixed'' asymptotic conditions can be obtained
from one another by duality transformation, which interchanges the
order and disorder fields $\sigma(x) \leftrightarrow \mu(x)$ and
reverses the sign of the energy field $\varepsilon(x)\mapsto
-\varepsilon(x)$. In the following, we will assume ``fixed''
asymptotic condition and $mR>-\h$.

Our main results are expressions for the order and disorder
correlation functions $\bra\sigma(x)\sigma(x')\ket$ and
$\bra\mu(x)\mu(x')\ket$, which admit a description in terms of a
Painlev\'e VI transcendent $w(\eta)$,
\eqn\painleve{\eqalign{
w'' = {1\over2}\({1\over w}+{1\over w-1}+{1\over w-\eta}\)w'^2
&-\({1\over\eta}+{1\over\eta-1}+{1\over w-\eta}\)w'
\cr
&+ \(\h-2\,\mR^2\){w(w-1)\over\eta(\eta-1)(w-\eta)}~.
}}
Here, primes denote derivative with respect to the projective
invariant $\eta$, related to the geodesic distance by
\eqn\projinvdist{
    \eta = \tanh^2\(\frc{\dist(x,x')}{2R}\),
}
and we have introduced the notation
\eqn\mudef{
    \mR = mR~.
}
From the usual parametrization in terms of auxiliary functions
$\chi(\eta)$ and $\varphi(\eta)$,
\eqn\param{\eqalign{
\(2R\)^{1/4}\,\bra\,\sigma(x)\,\sigma(x')\,\ket
&= e^{\chi(\eta)/2}\cosh(\varphi(\eta)/2)~,
\cr
\(2R\)^{1/4}\,\bra\,\mu(x)\,\mu(x')\,\ket
&= e^{\chi(\eta)/2}\sinh(\varphi(\eta)/2)~,
}}
we have found
\eqn\phichieqintro{\eqalign{
&\cosh^2 \varphi  = \frc1w ~,
\cr
&\chi' = {\eta(\eta-1)\over4w(w-\eta)(w-1)}w'^2
- {\eta\over2w(w-\eta)}w'
+\({1\over4} - r^2\){w-1\over(\eta-1)(w-\eta)}~.
}}
As expected, the Painlev\'e transcendent involved in this
description of two-point correlation functions is the same as that
involved in the description of the tau function of the Dirac
operator on the Poincar\'e disk (when specialized to appropriate
monodromies), found in Ref.~\Tracy\ and studied in Ref. \DoyonI.
However, it is a complicated matter to derive the former
description from the latter. We used very different and simpler
methods, following ideas developed in Ref. \FonsecaZamoII.

The appropriate solution to the Painlev\'e equation \painleve\ can
be fixed, for instance, by the short distance $\eta\to0$ behaviour
\eqn\wsmall{
    w = \mR^2\eta \ln^2\(k(r)^2\,\eta\) + O\(\eta^2\ln^4\eta\)~.
}
The constant $k(r)$, given by
\eqn\logk{
    \ln k(r) = \psi(\mR) + {1\over2\,\mR} + \gamma -\ln4
}
($\psi(x) = d\ln\Gamma(x)/dx$ and $\gamma$ is Euler's constant),
can be obtained from the vacuum expectation value of the energy
field $\bra\varepsilon\ket$ by applying conformal perturbation
theory  (see Appendix A). The power law in \wsmall, as well as the
behaviour $\chi(\eta) = \frc14\ln(\eta) + O(\eta^{\frc12})$ fixing
the integration constant for $\chi(\eta)$ in
\param, \phichieqintro, are specified by the leading short distance
behaviours
\eqn\ordernorm{
\dist(x,x')^{1/4}\bra\sigma(x)\sigma(x')\ket\ra 1~, \indent
\dist(x,x')^{1/4}\bra\mu(x)\mu(x')\ket\ra 1 \indent\hbox{as}\quad
\dist(x,x')\ra 0~.
}

This solution has the property that at large distances $\eta\to1$
it behaves as
\eqn\wlarge{
    1-w =  A(r)^2(1-\eta)^{1+2r}
+ O\((1-\eta)^{2+2\mR},(1-\eta)^{2+4\mR}\)~,
}
with the coefficient
\eqn\coefflarge{
    A(r) = {\Gamma\(\th+\mR\)\over4^{\mR} \sqrt\pi\,\Gamma(1+\mR)}~.
}
Furthermore, $\chi(\eta)$ approaches the constant $4\ln\bar s$, related to
the magnetization \refs{\DoyonI, \DoyonII}
\eqn\magnetization{
\bra\sigma\ket = (2R)^{-1/8}\,\bar s
\indent\hbox{with}\indent
    \bar s^2 = \sqrt2\,\prod_{n=1}^{\infty}
    \(1-{1\over4(\mR+n)^2}\over1-{1\over4 n^2}\)^n~.
}
The leading behaviour \wlarge\ is in agreement with the form
factors of the order and disorder fields which were calculated in
Ref.~\DoyonII\ as specialization of form factors of more general
fields in the Dirac theory on the Poincar\'e disk \DoyonI; in
Section~4 we verify these results in a simpler way using ideas of
Ref.~\FZamoI. The behaviour \wlarge\ also provides an alternative
description of the solution to the Painlev\'e equation \painleve\
describing the correlation functions. Together with \wsmall, it
gives a solution to the connection problem relating the
singular points $\eta=0$ and $\eta=1$.

Our results lay strong support, both analytical and numerical, for the
validity of the one-point function \magnetization\ and of the large
distance asymptotic of the two-point functions \param. We briefly analyzed
these quantities, along with the free energy, from the point of view of
the thermodynamics of the model. We found that, under natural assumptions
on an underlying lattice theory, there exist effective ``critical''
temperatures where the leading scaling behaviour changes drastically, with
``critical'' exponents closely related to those of mean field theory.

The paper is organized as follows. In Section~2, some technical details
regarding the field content and symmetries of the model \action\ are
presented and a space of asymptotic states is introduced. The algebra of
local integrals of motion of the doubled model are introduced in Section~3
and are then used in Section~4 to derive the non-linear differential
equations satisfied by the two-point correlation functions
$\bra\sigma(x)\sigma(x')\ket$ and $\bra\mu(x)\mu(x')\ket$, as well as the
associated form factors. Lastly, in Section~5, we discuss some
thermodynamical quantities in the Ising model on the pseudosphere.

\newsec{Ising Field Theory on a Pseudosphere}

The Ising field theory on the pseudosphere \action\ can be written
in terms of a free massive Majorana fermion $(\psi,\bar\psi)$ as
\eqn\IFTb{
{\cal A} = {1\over2\pi}\int_{|z|<1} d^2x~
\[\psi\bar\partial\psi+\bar\psi\partial\bar\psi
+{2i\,r\over1-z\bar z}\bar\psi\psi\]~,
}
where we have used, for the volume element $d^2x\, e^{\phi(x)}$ on
the Poincar\'e disk,
\eqn\metric{
e^{\phi(x)} = {4R^2\over\(1-z\bar z\)^2}~,
}
and for the energy field,
\eqn\energy{
\varepsilon(x) = i(2R)^{-1}(1-z\bar z)\,(\psi\bar\psi)(x)~.
}
In \IFTb, the parameter $r$ is related to the mass parameter $m$
and Gaussian curvature \curvature\ as in \mudef, $(z,\bar z)$ are
complex coordinates on the unit disk $|z|<1$, $\partial
\equiv\partial_z = \h(\partial_{\rm x} - i\partial_{\rm y})$,
$\bar\partial \equiv \partial_{\bar z} = \h(\partial_{\rm x} +
i\partial_{\rm y})$ and $d^2x\equiv d{\rm x}\,d{\rm y}$; $({\rm
x},{\rm y})$ are cartesian coordinates on the disk related in the
usual way to the complex coordinates, $z={\rm x} + i{\rm y}$,
$\bar z = {\rm x} - i{\rm y}$. In terms of complex coordinates,
the geodesic distance $\dist(x,x')$ between $x$ and $x'$ is given
by
\eqn\geodist{
\tanh^2{\dist(x,x')\over2R} = {(z-z')(\bar z-\bar z')
\over(1-z\bar z')(1-\bar z z')}~.
}

The chiral components $\psi$ and $\bar\psi$ obey the linear field
equations
\eqn\eqmotion{
\bar\partial\,\psi(x) = {i\,r\over1-z\bar z}\,\bar\psi(x)~,
\indent
\partial\,\bar\psi(x) = {-i\,r\over1-z\bar z}\,\psi(x)~,
}
and are normalized in \IFTb\ in accordance with the short-distance limit
\eqn\psinorm{
(z-z')\,\psi(x)\psi(x')\to 1~,
\indent
(\bar z-\bar z')\,\bar\psi(x)\bar\psi(x')\to 1
\indent\hbox{as}\indent |x-x'|\to0~.
}

The zero curvature limit $R\ra\infty$ corresponds to the familiar
theory of free massive Majorana fermion in flat space (and mass
$m$) after rescaling
\eqn\flat{
z\mapsto z/(2R)
\indent\hbox{and}\indent
\psi(x)\mapsto(2R)^{1/2}\,\psi(x)~,
}
with similar rescaling for $\bar z$ and $\bar\psi$.

\subsec{Symmetries}

The Ising field theory~\IFTb\ possesses an $SU(1,1)$ symmetry
group induced by the isometry group of the metric \metric. In particular,
the action is invariant under the coordinate transformation
\eqn\transf{
    z\mapsto z' = f(z) =  {az+\bar b\over b z+ \bar a}~, \indent \bar z
\mapsto
    \bar z' = \bar f(\bar z) = {\bar a \bar z+ b\over \bar b \bar z+ a}~,
}
where we can choose $a\bar a-b\bar b = 1$. Under this
transformation, the Fermi fields transform as
\eqn\fermitransf{
\psi(x) \mapsto \psi' (x') = (\partial f)^{-1/2}\, \psi(x)~,
\indent \bar\psi(x) \mapsto \bar\psi' (x') = (\bar\partial\bar
f)^{-1/2}\, \bar\psi(x)~.
}
In order for the full quantum theory to be $SU(1,1)$-invariant, we
have to impose appropriate $SU(1,1)$-invariant asymptotic
conditions at the disk boundary $|z|\to 1$, as will be discussed
later in this section.

It will be convenient to classify the local fields in the theory \IFTb\ by
their properties under $SU(1,1)$ transformations. A local field $\Or(x)$
will be said to have $SU(1,1)$-dimension $(h,\bar h)$ if it transforms
under \transf\ as
\eqn\Otransf{
\Or(x) \mapsto \Or' (x')
= (\partial f)^{-h} (\bar\partial\bar f)^{-\b{h}}\, \Or(x)~.
}
Thus, Fermi fields $\psi$ and $\b\psi$ have dimensions $\(\h,0\)$
and $\(0,\h\)$, respectively, and the energy field $\varepsilon$,
as defined in \energy, has dimension $\(0,0\)$.

One can construct local fields of higher $SU(1,1)$-dimension by using
the covariant derivatives
\eqn\covder{
    \cov\, \Or(x) = \(
    \partial-\frc{2h\b{z}}{1-z\b{z}}\) \Or(x)~, \indent
    \b\cov\, \Or(x) = \(
     \bar\partial  - \frc{2\b{h} z}{1-z\b{z}}\) \Or(x) ~,
}
where the holomorphic covariant derivative $\cov$ takes a field of
$SU(1,1)$-dimension $(h,\bar h)$ to a field of dimension
$(h+1,\bar h)$, and the anti-holomorphic covariant derivative
$\bar\cov$ to a field of dimension $(h,\bar h+1)$.

A basis for the isometry algebra can be taken as the generators of the
coordinates transformations
\eqn\transfocoord{
    z \mapsto z + \epsilon\, (1-z^2)~,
\indent
    z \mapsto z + i\,\epsilon\, (1+z^2)~,
\indent
    z \mapsto z + i\,\epsilon\, z~,
}
with conjugate transformations for $\bar z$ and where $\epsilon$
is a real infinitesimal parameter. These give rise to Lie
derivatives on the local fields~\Otransf, denoted respectively by
${\cal P}_\rx$, ${\cal P}_\ry$ and ${\cal R}$. Introducing the
notation \hbox{${\cal P} = \h ({\cal P}_\rx - i\,{\cal P}_\ry)$},
${\b{\cal P}} = \h ({\cal P}_\rx + i\,{\cal P}_\ry)$, they are
given by
\eqn\Lie{\eqalign{
{\cal P}\,\Or(x) &= \(\partial-\bar z^2\bar\partial - 2\bar h\bar
z\)\Or(x)~,
\cr
{\bar{\cal P}}\,\Or(x) &= \(\bar\partial- z^2\partial - 2 h z\)\Or(x)~,
\cr
{\cal R}\, \Or(x) &= i\(z\partial- \bar z\bar\partial + h - \bar
h\)\Or(x)~.
}}
The conservation of the associated currents,
\eqn\check{
\bar\partial\(\psi\,{\cal P}\psi\) = \partial\(\bar\psi\,{\cal
P}\bar\psi\)\,, \indent \bar\partial\(\psi\,\bar{\cal P}\psi\) =
\partial\(\bar\psi\,\bar{\cal P}\bar\psi\)\,, \indent
\bar\partial\(\psi\,{\cal R}\psi\) = \partial\(\bar\psi\,{\cal
R}\bar\psi\)\,,
}
is a simple consequence of the field equations \eqmotion. In
general, one can construct conserved currents by applying an
arbitrary number of Lie derivatives on the Fermi fields, i.e.
\eqn\othercurr{
\bar\partial\({\cal L}'\psi~{\cal L}\psi\) = \partial\({\cal
L}'\bar\psi~{\cal L}\bar\psi\)~,
}
where ${\cal L}' = {\cal P}^{l_1}\,\bar{\cal P}^{l_2}\,{\cal
R}^{l_3}$, ${\cal L} = {\cal P}^{n_1}\,\bar{\cal P}^{n_2}\,{\cal
R}^{n_3}$, $l$'s, $n$'s being some non-negative integers.

As usual, integrals of motion can be written as integrals of the
currents over appropriate lines on the pseudosphere. It will
be convenient in what follows to choose for these lines a particular
family of geodesics on the pseudosphere parametrized by elements of a
non-compact subgroup of the $SU(1,1)$ isometry group. More
precisely, consider the system of ``isometric'' coordinates
$(\xi,\b\xi)$ related to coordinates $(z,\bar z)$ on the
Poincar\'e disk by
\eqn\gcoord{
z = \tan\xi~, \indent \bar z = \tan\bar\xi~,
}
where $\xi = \xi_\rx + i\, \xi_\ry$ and $\bar\xi = \xi_\rx -
i\,\xi_\ry$ (see Figure~1).

\fig{1}{In the $\xi$-plane, related to the unit disk by the map \gcoord, the
pseudosphere lies on the strip $-\pi/4<\xi_\rx<\pi/4$ and
$-\infty<\xi_\ry<\infty$; the lines $\xi_{\rm y} = {\rm const.}$
correspond to a particular class of geodesics.}
{\epsfysize=3.8cm\epsfbox{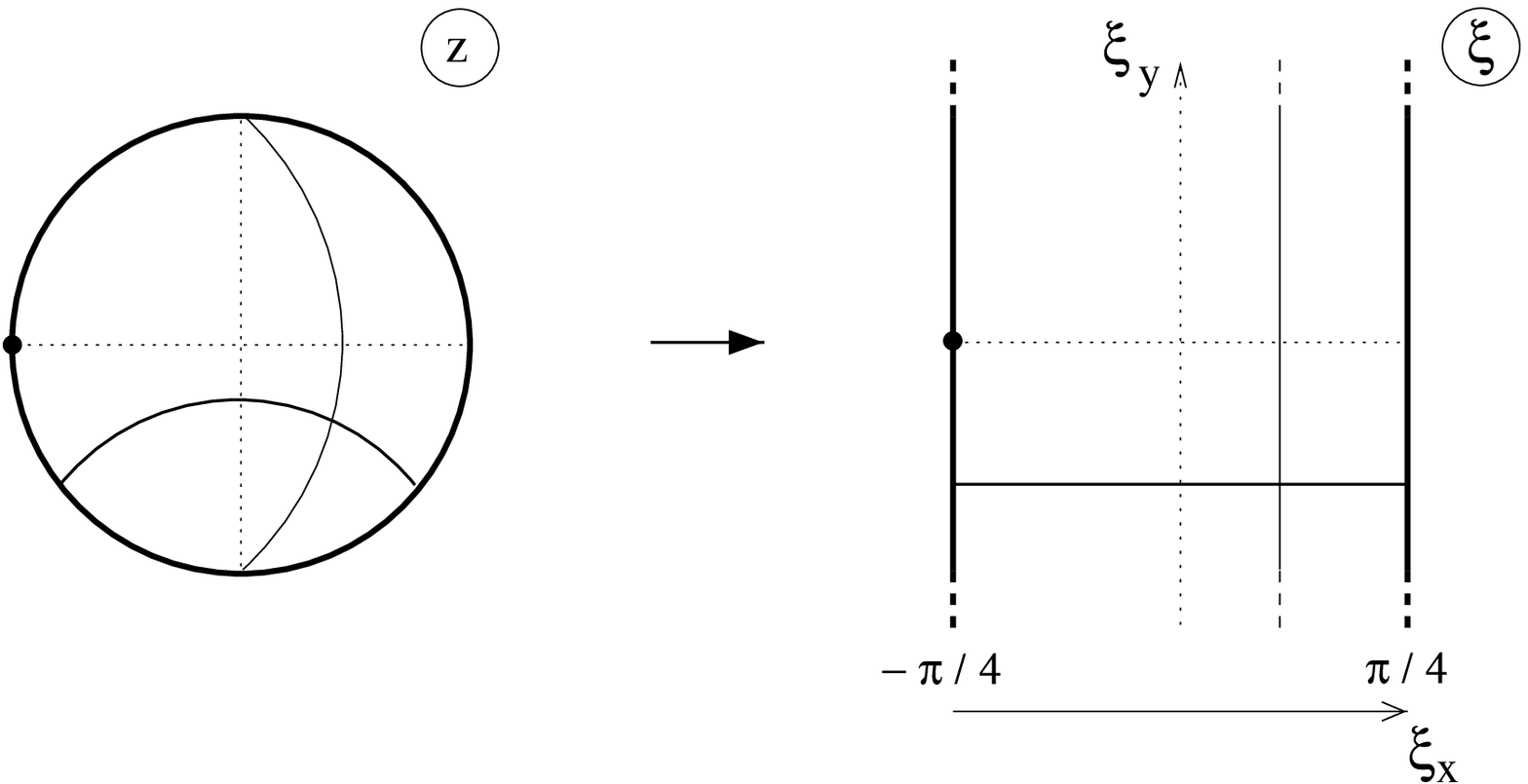}}

The transformations
\eqn\transfocoordxi{
    \xi_\ry\mapsto \xi_\ry+q~,
\indent\hbox{for arbitrary real $q$}~,
}
are the integrated versions of the second infinitesimal
transformation of \transfocoord\ and form a non-compact subgroup
of $SU(1,1)$. It will be convenient to write the integrals of
motion associated to the conserved currents \check\ as integrals
over lines of constant $\xi_\ry$,
\eqn\charges{\eqalign{
{\bf P} = {1\over4\pi}\int_{\iline} &\Big[\psi\,{\cal P}\psi\, dz
- \bar\psi\,{\cal P}\bar\psi\, d\bar z \Big]~,
\cr
{\bf \bar P} =
{1\over4\pi}\int_{\iline} &\Big[\bar\psi\,\bar{\cal P}\bar\psi\,
d\bar z - \psi\,\bar{\cal P}\psi\, d z \Big]~,
\cr
{\bf R} =
{1\over4\pi i}\int_\iline &\Big[\psi\,{\cal R}\psi\, dz -
\bar\psi\,{\cal R}\bar\psi\, d\bar z \Big]~,
}}
where
\eqn\qline{
    \iline = \{(z,\bar z)\; | ~
        -\pi/4 < \xi_\rx <\pi/4 ~,~~ \xi_{\rm y}={\rm const.}\}~.
}
Inside correlation functions with some local fields, these
integrals are independent of the value of $\xi_\ry$ associated to
the path $\iline$, except for contributions at the positions of
the fields, given by the corresponding Lie derivatives,
\eqn\chargeaction{
[{\bf P},{\cal O}(x)] = i{\cal P}\, {\cal O}(x)~,
\indent
[{\bf \bar P},{\cal O}(x)] = -i{\bar{\cal P}}\, {\cal O}(x)~,
\indent
[{\bf R},{\cal O}(x)] = {\cal R}\, {\cal O}(x)~,
}
where, for instance,
\eqn\egcomm{
[{\bf P},{\cal O}(x_0)] = -{1\over4\pi}\oint_{x_0}
\Big[\psi\,{\cal P}\psi\, dz - \bar\psi\,{\cal P}\bar\psi\, d\bar
z \Big] {\cal O}(x_0)~,
}
the contour being taken around the point $x_0$ in
counter-clockwise direction. The integrals of motion \charges\
satisfy the $su(1,1)$ algebra
\eqn\algebra{ [{\bf P}, {\bf R}] = -i\,{\bf P}~,
\indent [{\bf \bar P}, {\bf R}] = i\,{\bf \bar P}~, \indent [{\bf
P}, {\bf \bar P}] = -2i\,{\bf R}~,
}
and the field equations \eqmotion\ specify the Casimir of the
representation generated by the Fermi fields,
\eqn\Casimir{
\h\Big( [{\bf P}, [{\bf \bar P},\psi]]
+ [{\bf \bar P}, [{\bf P},\psi]]\Big)
- [{\bf R},[{\bf R},\psi]] = \(\mR^2 - {1\over4}\) \psi~,
}
with a similar equation for $\b\psi$.

\subsec{Hilbert Space}

Here, we present a space of asymptotic states for the Ising field
theory \IFTb. Details regarding its construction can be found in
Appendix B.

A particularly convenient Hilbert space is the one obtained by
associating time translations to isometry transformations
\transfocoordxi, so that $\xi_\rx$ represents the space coordinate
and $\xi_\ry$ the Euclidean time coordinate. A basis can be
obtained by diagonalizing the corresponding Hamiltonian
\eqn\Hcharge{
    {\bf H} = {\bf P} + {\bf \bar P}~.
}
As we will see, the corresponding states can be interpreted as
``particle states'', since they form an irreducible representation
of the $SU(1,1)$ symmetry group.

The Hilbert space is defined as a module for the canonical
equal-time anti-commutation relations of the fermion operators
$\psiiso(\xi_\rx,\xi_\ry)$, $\bpsiiso(\xi_\rx,\xi_\ry)$ in the isometric
system of coordinates\foot{The Fermi fields in isometric coordinates are
related to the Fermi fields on the Poincar\'e disk by
\fermitransf, that is, $\psiiso(\xi_\rx,\xi_\ry) = \psi(x)/\cos(\xi)
,\,\bpsiiso(\xi_\rx,\xi_\ry) = \bar\psi(x)/\cos(\bar\xi)$.}:
\eqn\canonical{\eqalign{
    \{\psiiso(\xi_\rx,\xi_\ry),\,\psiiso(\xi_\rx',\xi_\ry)\} &=
    -2\pi i\,\delta(\xi_\rx-\xi_\rx')~,\cr
    \{\bpsiiso(\xi_\rx,\xi_\ry),\,\bpsiiso(\xi_\rx',\xi_\ry)\} &=~~
    2\pi i\,\delta(\xi_\rx-\xi_\rx')~.
}}
Invariance under the subgroup described by \transfocoordxi\
imposes that the Fermi fields vanish as $\xi_\ry \to \pm \infty$,
which gives the following conditions on the vacuum state:
\eqn\vaccond{
    \lim_{\xi_\ry\to-\infty} \psiiso(\xi_\rx,\xi_\ry)|\vac\> = 0~,
\indent\indent
    \lim_{\xi_\ry\to+\infty} \<\vac|\psiiso(\xi_\rx,\xi_\ry) = 0~,
}
with similar conditions for $\bpsiiso$. Correlation functions of
local fields are then expressed as time-ordered vacuum expectation
values of corresponding operators; the time ordering puts
operators from left to right in decreasing values of their
variable $\xi_\ry$.

The Hilbert space is further specified by the asymptotic
conditions imposed on the Fermi fields at the disk boundary. In
order to guarantee their stability, one has to choose such
conditions giving a Hilbert space on which the Hamiltonian has
real eigenvalues bounded from below. Among these, we shall only
consider those which respect the $SU(1,1)$ symmetry. We note
however that for $-\h< \mR< \h$ there are stable asymptotic
conditions which are not $SU(1,1)$ invariant; we intend to report
on these regimes in a future publication.

For $r>\h$, finiteness of matrix elements of the Hamiltonian gives
the asymptotic conditions imposing Fermi fields to vanish as $\sim
e^{-m\,\dist}$ when $\dist$, the geodesic distance between the
origin, say, and the position of the Fermi field, goes to
infinity. These asymptotic conditions correspond to ``fixed''
asymptotic conditions on the order field $\sigma$. For $0<\mR<\h$,
the above asymptotic conditions are also stable, and there is an
additional set of stable asymptotic conditions, by which the Fermi
fields diverge as $\sim e^{m\dist}\, (1+O(e^{-\dist/R}))$ as the
geodesic distance to the origin goes to infinity. This second set
corresponds to ``free'' asymptotic conditions on the order field
$\sigma$. The field theory with this second set of asymptotic
conditions can be obtained by analytically continuing the field
theory with the first set of asymptotic conditions from the region
$0<\mR<\h$ to the region $-\h<\mR<0$, and vice versa. Hence it
will be sufficient in what follows to consider only the conditions
specifying the asymptotic behaviour $\sim e^{-m\,\dist}$ of the
Fermi fields for $\mR>0$.

With these asymptotic conditions, the Fermi fields, obeying
Eqs.~\eqmotion, admit expansions in partial waves as
\eqn\waves{\eqalign{
\psi(\xi_\rx,\xi_\ry)
&= \cos\xi\sum_{n=0}^\infty
\Big[e^{i{\pi\over2}n}\,A^\dagger_n e^{\omega_n\xi_\ry} g_n(\xi_\rx)
- i\,e^{-i{\pi\over2} n}\, A_n e^{-\omega_n\xi_\ry} \bar
g_n(\xi_\rx)\Big]~,
\cr
\bar\psi(\xi_\rx, \xi_\ry)
&=\cos\bar\xi\sum_{n=0}^\infty
\Big[
e^{i{\pi\over2}n} A^\dagger_n e^{\omega_n\xi_\ry}\bar g_n(\xi_\rx)
+ i\,e^{-i{\pi\over2} n}\, A_n e^{-\omega_n\xi_\ry} g_n(\xi_\rx)\Big]~,
}}
with discrete energy spectrum
\eqn\Heigenvalue{
\omega_n = 2\mR + 2n + 1 \indent (n\ge0)~.
}
Partial waves are given by
\eqn\gfunction{
g_n(\xi_\rx) = {2^{1-2\mR}\,\sqrt{\pi}\over\Gamma\(\mR+\th\)}
{\Gamma(2\mR+n+1)^{1/2}\over\Gamma(n+1)^{1/2}}
e^{-i\omega_n\xi_{\rx}-i{\pi\over2}n }\, \(1+e^{4i\xi_\rx}\)^\mR
F\(-n,\mR;1+2\mR;1+e^{4i\xi_\rx}\),
}
with $\bar g_n(\xi_\rx)$ denoting its complex conjugate, and
$F(a,b;c;x)$ stands for the Gauss hypergeometric function, here
specialized to polynomials; phases in the decomposition \waves\
were chosen for later convenience when introducing particle
states. The functions $g_n(\xi_{\rm x})$, $\bar g_n(\xi_{\rm x})$
satisfy the orthogonality relations
\eqn\orto{\eqalign{
\int_{-\pi/4}^{\pi/4} d\xi_\rx
\Big[g_n(\xi_\rx)\bar g_{n'}(\xi_\rx)
+ \bar g_n(\xi_\rx)g_{n'}(\xi_\rx)\Big]
&= 4\pi\,\delta_{n,n'}~,
\cr
\int_{-\pi/4}^{\pi/4} d\xi_\rx
\Big[g_n(\xi_\rx) g_{n'}(\xi_\rx)
- \bar g_n(\xi_\rx) \bar g_{n'}(\xi_\rx)\Big]
&= 0~,
}}
as well as the relations
\eqn\gduality{
g_n(\xi_{\rm x})\Big|_{r=-\h}
= -i\,g_{n-1}(\xi_{\rm x})\Big|_{r=\h}~,
\indent
\bar g_n(\xi_{\rm x})\Big|_{r=-\h}
= i\,\bar g_{n-1}(\xi_{\rm x})\Big|_{r=\h}~.
}

The creation and annihilation operators $A^\dagger_n$ and $A_n$
($n\ge0$) in \waves\ satisfy canonical anti-commutation relations
as consequence of \canonical:
\eqn\commutatormodes{
    \{A^\dagger_n, A_{n'}\} =
    \delta_{n,n'}~,
\indent
\{A^\dagger_n,A^\dagger_{n'}\} =
    \{A_n,A_{n'}\} = 0~,
}
with the vacuum state $|{\rm vac}\ket$ obeying, from \vaccond,
\eqn\vaccond{
A_n|\vac\ket = 0\indent\hbox{for all}\indent n\ge0~.
}
A basis of $N$-particle states is obtained from the set of states
\eqn\particlestate{
|n_1\,\ldots\,n_N\ket
\equiv A^\dagger_{n_1}\,\ldots\, A^\dagger_{n_N}|{\rm vac}\ket~,
}
which diagonalize the Hamiltonian,
\eqn\Haction{
{\bf H}\,|n_1\,\ldots\, n_N\ket
= \(\sum_{i=1}^N \omega_{n_i}\) |n_1\,\ldots\, n_N\ket
}
with energy eigenvalues $\omega_n$, Eq.~\Heigenvalue. The
discretization of the energy spectrum is essentially a consequence
of requiring trivial monodromy of the hypergeometric functions
involved in the partial waves \gfunction\ as $\xi_\rx \to \xi_\rx
+ \pi/2$, a necessary condition in order to ensure the proper
vanishing asymptotic behaviour at the boundary of the disk.

The action of the operators $\P$, $\Pbar$ and $\R$, defined in
\charges, can be easily determined from the fact that the above
Hilbert space provides a lowest weight module for $SU(1,1)$. The
raising and lowering operators, ${\bf J}_+$ and ${\bf J}_-$
respectively, are given by
\eqn\Jcharges{
{\bf J}_\pm = {\bf P} - \bar {\bf P} \pm 2i\,{\bf R}~,
}
and are related by hermitian conjugation, ${\bf J}_+^\dagger = {\bf
J}_-$. Together with the Hamiltonian \Hcharge, they satisfy the algebra
\eqn\Halgebra{
[{\bf H},{\bf J}_\pm] = \pm 2\,{\bf J}_\pm~,
\indent
[{\bf J}_-, {\bf J}_+ ] = 4\,{\bf H}~,
}
from which the action of ${\bf J}_\pm$ on eigenstates of the
Hamiltonian follows:
\eqn\Jaction{
{\bf J}_+|n\ket = \alpha_n |n+1\ket~,
\indent
{\bf J}_-|n\ket = \alpha_{n-1}|n-1\ket~,
}
with
\eqn\alphan{
\alpha_n =  2\sqrt{(n+1)(2\,\mR+n+1)}~.
}
We thus have the action of the rotation operator ${\bf R}$,
\eqn\Rstate{
4i\,{\bf R}|n\ket = \alpha_n|n+1\ket - \alpha_{n-1}|n-1\ket~,
}
that we will use in Section~4 when deriving form factors.

\subsec{Local Fields}

Besides the Fermi and the energy fields, other local fields are
present in the theory. Two spin fields associated to the ${\IZ}_2$
symmetry $(\psi,\bar\psi)\mapsto(-\psi,-\b\psi)$ of the action
\IFTb\ can be defined, the order field $\sigma(x)$ and the
disorder field $\mu(x)$. They are not mutually local with respect
to the Fermi fields, since the products
\eqn\products{
    \psi(x)\sigma(x')~,\quad \b\psi(x)\sigma(x')~, \quad
    \psi(x)\mu(x')~,\quad \b\psi(x)\mu(x')
}
acquire negative signs when the point $x$ is brought around $x'$.
This property does not define the fields $\sigma$ and $\mu$
uniquely. Besides having this property, they are required to be
``primary'' with respect to the action of the Fermi fields in the
operator algebra. This fixes operator product expansions (OPE) of
the form
\eqn\OPE{\eqalign{
\psi(x)\sigma(x') = \sum_{n=0}^\infty \;c_n\;
\Big[~~&\sqrt{\frc{i}2}\,s_n(x,x') \,u_n(\eta)\,\cov^n\mu(x') +
\sqrt{\frc{-i}2}\,\bar t_n(x,x')\,v_n(\eta)\,\bar\cov^n\mu(x')
\Big]~,
    \cr
\bar\psi(x)\sigma(x') = \sum_{n=0}^\infty \;c_n\;
\Big[-&\sqrt{\frc{i}2}\,t_n(x,x')\,v_n(\eta)\,\cov^n\mu(x') +
\sqrt{\frc{-i}2}\,\bar s_n(x,x')\,u_n(\eta)\,\bar\cov^n\mu(x')
\Big]~.
}}
The factors $s_n$ and $t_n$ are given by
\eqn\factorsOPE{\eqalign{
    s_n(x,x') &= \(\frc{1-z'\b{z}'}{1-z\b{z}'}\)^{n+\frc12} \,
    (z-z')^{n-\frc12}~,
\cr
    t_n(x,x') &= \(\frc{1-z'\b{z}'}{1-z\b{z}'}\)^{n+\frc12}
    \frc{(z-z')^{n+\frc12}}{1-\b{z}z'}~;
}}
$s_n$ transforms under the representation
$\(\th,0\)$ in $x$ and $\(-n,0\)$ in $x'$, and $t_n$ under $\(0,\th\)$ in 
$x$ and $\(-n,0\)$ in $x'$; the factors $\bar
s_n$, $\bar t_n$ are their complex conjugates; the projective
invariant $\eta$ is given by \projinvdist, \geodist, i.e.
\eqn\projinv{
\eta = {(z-z')(\bar z-\bar z')
\over(1-z\bar z')(1-\bar z z')}~;
}
the functions
$u_n(\eta)$ and $v_n(\eta)$,
\eqn\ggbar{\eqalign{
u_n(\eta) &= (1-\eta)^{\mR}\, F\(\mR,\mR+\th+n;\h+n;\eta\)~,
\cr
v_n(\eta) &= \frc{i\mR}{n+1/2}(1-\eta)^{\mR}\,
F\(\mR+1,\mR+\th+n;\frc32+n;\eta\)~,
}}
are determined by the field equations \eqmotion; and ${\cal D}$,
$\bar{\cal D}$ are the covariant derivatives introduced
in \covder.

The constants $c_n$ can be determined, say, from requiring associativity
of the operator algebra on $\psi(x)\psi(x')\partial\sigma(0)$,
\eqn\constOPECFT{
c_0 = 1~,
\indent
c_n = 2\,/\(\th\)_n\indent(n\ge1)~,
}
with $\(\frc12\)_n = \Gamma\(\th+n\)/\Gamma\(\th\)$.

There are similar expressions for the products $\psi(x)\mu(x')$,
$\bar\psi(x)\mu(x')$, obtained from the above OPE \OPE\ by interchanging
$\sigma\leftrightarrow\mu$ and $\sqrt{i}\leftrightarrow \sqrt{-i}$.
These completely define the fields $\sigma$ and $\mu$, together with the
normalization \ordernorm\ and the convention that they are $SU(1,1)$
scalars (so that their $SU(1,1)$-dimensions are $h= \bar h = 0$).

\newsec{Conserved Charges of the Doubled Ising Field Theory}

In order to derive Ward identities for correlation functions and
form factors, we shall follow closely the method presented in
Ref.~\FonsecaZamoII. In particular, we start by introducing a
system of two identical copies of the Ising field theory \IFTb,
here referred to as ``copy $a$'' and ``copy $b$'', a trick known
to simplify many aspects of the Ising theory
\refs{\Luther,\Drouffe}. The fields $\psi_a$, $\bar\psi_a$,
$\sigma_a$, $\mu_a$ will denote respectively the two fermionic and
the order and disorder fields belonging to copy $a$, while
$\psi_b$, $\bar\psi_b$, $\sigma_b$, $\mu_b$ will denote those
belonging to copy $b$. In addition, it is convenient to require
Fermi fields from different copies to anti-commute. This can be
achieved by introducing two auxiliary Klein factors, $\eta_a$ and
$\eta_b$, defined by
\eqn\klein{
\eta_a^2 = 1~,\quad
\eta_b^2 = 1~,\quad
\eta_a\,\eta_b = -\eta_b\,\eta_a~,
}
and modifying original fields according to
\eqn\newfields{
\psi_a(x)\mapsto\eta_a\,\psi_a(x)~,\quad
\bar\psi_a(x)\mapsto\eta_a\,\bar\psi_a(x)~,\quad
\sigma_a(x)\mapsto\sigma_a(x)~,\quad
\mu_a(x)\mapsto\eta_a\,\mu_a(x)~,
}
with similar redefinition for copy $b$. These redefinitions do not
change any of the correlation functions involving only fields
belonging to a given copy, as long as we assume ``fixed''
asymptotic condition, so that $\bra\mu\ket=0$. Correlation
functions involving fields belonging to both copies factorize into
correlation functions in the model with a single copy in a simple
manner. For instance,
$\bra\sigma_a(x)\sigma_b(x)\,\sigma_a(x')\sigma_b(x')\ket =
\bra\sigma(x)\sigma(x')\ket^2$ and
$\bra\mu_a(x)\mu_b(x)\,\mu_a(x')\mu_b(x')\ket = -
\bra\mu(x)\mu(x')\ket^2$, the minus sign in the latter coming from
the redefinitions \newfields.

This doubled model is nothing else than a theory of
free Dirac field on a pseudosphere. The $U(1)$ invariance of the
Dirac theory is translated into an invariance under ``rotations''
among the two copies, with the associated charge being
\eqn\Qcharge{
{\bf Z}_0 = {1\over2\pi}\int_{\iline} \Big[\psi_a\psi_b\, dz -
\bar\psi_a\bar\psi_b\,d\bar z\Big]~,
}
where the contour is taken on an equal-time slice, given by
\qline. It acts on the Fermi fields as
\eqn\Qchargefermi{
[{\bf Z}_0, \psi_a(x)] = i\psi_b(x)~, \indent [{\bf Z}_0,
\psi_b(x)] = -i\psi_a(x)~.
}
with similar relations for $\bar\psi$'s.

Additional local integrals of motion for the doubled model can be directly
obtained from the charges $\P$, $\Pbar$ and $\R$, defined in \charges, of
each individual copy,
\eqn\chargesX{
\X_1 = \P_a - \P_b~,
\indent
\X_{-1} = \Pbar_a - \Pbar_b~,
\indent
\X_0 = \R_a - \R_b~,
}
and from the commutators of these with the $U(1)$ charge \Qcharge, such as
\eqn\chargesY{
\Y_1 = {i\over2}\, [\X_1,{\bf Z}_0]~,
\indent
\Y_{-1} = {i\over2}\,[\X_{-1},{\bf Z}_0]~,
\indent
\Y_0 = {i\over2}\, [\X_0,{\bf Z}_0]~.
}
In terms of integrals of local currents,
\eqn\chargesYint{\eqalign{
{\bf Y}_1 &= {1\over2\pi}\int_{\iline}
\(\psi_a{\cal P}\psi_b\, dz
- \bar\psi_a{\cal P}\bar\psi_b\, d\bar z\)~,
\cr
{\bf Y}_{-1} &= {1\over2\pi}\int_{\iline}
\(\bar\psi_a{\bar{\cal  P}}\bar\psi_b\, d\bar z
- \psi_a{\bar{\cal P}}\psi_b\, dz\)~,
\cr
{\bf Y}_0 &=
{1\over2\pi i}\int_{\iline} \( \psi_a{\cal  R}\psi_b\, d z
- \bar\psi_a{\cal  R}\bar\psi_b\, d\bar z\)~.
}}
Yet other integrals of motion obtained from the above are:
\eqn\chargesZ{\eqalign{
&{\bf Z}_2 = -{i\over2}\, [{\bf X}_1,{\bf Y}_1]~,
\indent\hskip66pt
{\bf Z}_{-2} = {i\over2}\, [{\bf X}_1,{\bf Y}_{-1}]~,
\cr
&{\bf Z}_1 = -{i\over2}\, [{\bf X}_1,{\bf Y}_0]
= -{i\over2}\, [{\bf X}_0,{\bf Y}_1]~,
\indent
{\bf Z}_{-1} = {i\over2}\, [{\bf X}_{-1},{\bf Y}_0]
= {i\over2}\, [{\bf X}_0,{\bf Y}_{-1}]~,
\cr
&{\bf\widetilde Z}_0 = -{i\over2}\, [{\bf X}_0,{\bf Y}_0]~.
}}
All these can be straighforwardly written as integrals of local
densities, quadratic in the Fermi fields and their Lie
derivatives. We present here the expressions for ${\bf Z}_2$ and
${\bf Z}_{-2}$,
\eqn\chargesQtwoint{\eqalign{
{\bf Z}_2 &= {1\over2\pi}\int_{\iline}
\({\cal P}\psi_a{\cal P}\psi_b\, dz
- {\cal P}\bar\psi_a{\cal P}\bar\psi_b\, d\bar z\)~,
\cr
{\bf Z}_{-2} &= {1\over2\pi}\int_{\iline}
\(\bar{\cal P}\bar\psi_a{\bar{\cal  P}}\bar\psi\, d\bar z
- \bar{\cal P}\psi_a{\bar{\cal P}}\psi_b\, dz\)~.
}}

Some additional commutation relations among these integrals of
motion are\foot{In the zero curvature limit $R\ra\infty$ the algebra
specified by \chargesY, \chargesalgebra\ corresponds to the well known
$\widehat{sl}(2)$ affine algebra of level zero \Leclair\ after appropriate
rescaling dictated by \flat; otherwise, the symmetry algebra of the
doubled model fails to qualify as a Kac-Moody algebra due to non-trivial
commutation relations of ${\bf\widetilde Z}_0$, in Eq.~\chargesZ, with
other charges.},
\eqn\chargesalgebra{\eqalign{
& [{\bf Y}_1,{\bf Z}_0] = 2i\,{\bf X}_1~,
\quad
[{\bf Y}_{-1},{\bf Z}_0] = 2i\,{\bf X}_{-1}~,
\quad
[{\bf Y}_0,{\bf Z}_0] = 2i\,{\bf X}_0~,
\cr
&[{\bf X}_1,{\bf Y}_{-1}] = [{\bf X}_{-1},{\bf Y}_1]
= 2i\(\mR^2-{1\over4}\){\bf Z}_0
+ 2i\,{\bf\widetilde Z}_0~.
}}
A straightforward way to obtain these relations is to calculate
them on the module provided by the Fermi fields (and their
descendants) of both copies, since on them the action of the
charge \Qcharge\ is simple, and to use the equations of motion in
the form \Casimir.

The OPE's of Fermi and spin fields can be used to calculate the
actions of the conserved charges on the order and disorder fields.
A particularly simple case is when the fields are located at the
center of the unit disk. For the operator ${\bf Z}_0$, we have the
commutators
\eqn\Qaction{\eqalign{
&[{\bf Z}_0,\sigma_a(0)\,\sigma_b(0)] = 0~,
\cr
&[{\bf Z}_0,\sigma_a(0)\,\mu_b(0)] = -i\,\mu_a(0)\,\sigma_b(0)~,
}
\indent
\eqalign{
&[{\bf Z}_0,\mu_a(0)\,\mu_b(0)] = 0~,
\cr
&[{\bf Z}_0,\mu_a(0)\,\sigma_b(0)] = i\,\sigma_a(0)\,\mu_b(0)~.
}}
Another useful commutator is the one involving the combination
${\bf Y}_1 - {\bf Y}_{-1}$,
\eqn\Yaction{
[{\bf Y}_1 - {\bf Y}_{-1},\, \mu_a(0)\mu_b(0)]
= i\big[{\bf H}_a - {\bf H}_b,\, \sigma_a(0)\sigma_b(0)\big]~,
}
where ${\bf H}_a$, ${\bf H}_b$ are the Hamiltonians \Hcharge\ of each
individual copy.

The action at an arbitrary point can be obtained either directly from the
OPE \OPE\ or from applying appropriate $SU(1,1)$ transformations to their
action on fields at the center of the disk, as shown in Appendix C. We
shall need the results
\eqn\Qddsigmamu{\eqalign{
i\[{\bf Z}_0,\partial\bar\partial\sigma_a(x)\,\mu_b(x)\]
=\partial\mu_a(x)\,\bar\partial\sigma_b(x)
&+\bar\partial\mu_a(x)\,\partial\sigma_b(x)
-\partial\bar\partial\mu_a(x)\,\sigma_b(x)
\cr
&+ {\mR^2-1/4\over(1-z\bar z)^2}\,\mu_a(x)\,\sigma_b(x)~,
}}
and
\eqn\Qtwosigmasigma{\eqalign{
\[{\bf Z}_2,\sigma_a(x)\,\sigma_b(x)\]
&=2\partial\mu_a(x)\,\partial\mu_b(x)
-\partial^2\mu_a(x)\,\mu_b(x)
-\mu_a(x)\,\partial^2\mu_b(x)
\cr
&- \bar z^4\Big(2\bar\partial\mu_a(x)\,\bar\partial\mu_b(x)
-\bar\partial^2\mu_a(x)\,\mu_b(x)
-\mu_a(x)\,\bar\partial^2\mu_b(x)\Big)
\cr
&+ 2\bar z^3\Big(\bar\partial\mu_a(x)\,\mu_b(x)
+\mu_a(x)\,\bar\partial\mu_b(x)\Big)~,
}}
plus a similar commutator $\[{\bf Z}_2,\mu_a(x)\,\mu_b(x)\]$, given by
\Qtwosigmasigma\ with $\mu$ replaced by $\sigma$.

Modes $A_n,\,A_n^\dagger$ as introduced in \waves\ will
be considered to enter the partial wave expansion of
$\psi_a,\,\bar\psi_a$, whereas modes $B_n,\,B_n^\dagger$
with the same commutation relations (and anti-commuting with the
$A$'s) enter the partial wave expansion of $\psi_b,\,\bar\psi_b$
in the same manner. The Hilbert space of the doubled model is
isomorphic to the tensor product of two copies of the Hilbert
space described in Section 2.2, with particle states denoted by
\eqn\doublestates{
|n_1\,\ldots\, n_N;\, m_1\,\ldots\, m_M\ket\equiv
A^\dagger_{n_1}\cdots A^\dagger_{n_N} \,
B^\dagger_{m_1}\cdots B^\dagger_{m_M} |\vac\ket~.
}

The action of conserved charges of the doubled model on these
asymptotic states can be easily deduced from the action of the
conserved charges $\P,\,\Pbar,\,\R$ on the states described in
Section 2.2, along with
\eqn\actionQstates{
{\bf Z}_0|\vac\ket = 0 ~,
\indent
[{\bf Z}_0,A^\dagger_n] = iB^\dagger_n~,
\indent
[{\bf Z}_0,B^\dagger_n] = -iA^\dagger_n~,
}
which follows directly from the decomposition \waves\ and \Qcharge. The
fact that all local integrals of motion anihilate the vaccum state will
then give rise to useful Ward identities.

\newsec{Correlation Functions and Form Factors}

In this section we use the Ward identities associated with the symmetries
of the single and the doubled models in order to derive the differential
equations satisfied by the spin fields as well as the corresponding form
factors.

\subsec{The Correlation Functions}

The representation \painleve, \phichieqintro\ of the two-point functions
of order and disorder fields
\eqn\GGtilde{
    G(x,x') = \bra\,\sigma(x)\sigma(x')\,\ket~,\indent
    \tilde G(x,x') = \bra\,\mu(x)\mu(x')\,\ket~,
}
can be derived using the symmetries of the doubled model described in the
previous section. In particular, using \Qtwosigmasigma\ we can rewrite the
Ward identity \eqn\wardI{
\bra \vac|\,
\big[{\bf Z}_2,\, \sigma_a(x)\sigma_b(x)\,\mu_a(x')\mu_b(x')\big]
\,|\vac\ket = 0~,
}
as the quadratic differential equation
\eqn\eqI{\eqalign{
&~~
\partial \tilde G\,\partial \tilde G
- \partial^2 \tilde G\,\tilde G
- \bar z^4\(\bar\partial \tilde G\,\bar\partial \tilde G
- \bar\partial^2 \tilde G\,\tilde G\)
+ 2\bar z^3\bar\partial\tilde G\,\tilde G
\cr
=~&\partial' G\,\partial' G
- \partial'^2 G \, G
- \bar z'^4\(\bar\partial' G\,\bar\partial' G
- \bar\partial'^2 G\, G\)
+ 2\bar z'^3 \bar\partial' G\, G~,
}}
where $\partial=\partial_z$, $\partial'=\partial_{z'}$,
$\bar\partial=\partial_{\bar z}$ and $\bar\partial'=\partial_{\bar z'}$;
it is important to keep track of the Klein factors in the definition of
the fields $\mu_a$, $\mu_b$ in order to get the signs in the above
equation.

Also, using \Qddsigmamu, we find that the Ward identity \eqn\wardII{
\bra\vac|\, \big[{\bf Z}_0,\, \partial\bar\partial\sigma_a(x)\mu_b(x)\,
\mu_a(x')\sigma_b(x')\big]\,|\vac\ket = 0~,
}
gives
\eqn\eqII{
(1-z\bar z)^2
\[\partial\bar\partial \tilde G\, G
+ \tilde G\, \partial\bar\partial G
- \partial\tilde G\,\bar\partial G
- \bar\partial\tilde G\,\partial G\]
= \(\mR^2-{1\over4}\)\tilde G\,G~.
}

We are interested in a solution that respects the $SU(1,1)$ symmetry, in
which case the two-point correlation functions are simply functions of
the projective invariant \projinv. Then, Eqs.~\eqI\ and \eqII\ above
imply the set of equations\foot{Similar calculations can be easily
performed for the Ising field theory on a sphere, as described in the
Appendix~E.}
\eqna\eqsGGtilde$$\eqalignno{
&\eta(1-\eta)\(G' G' - G'' G + \tilde G'\tilde G'-\tilde G''\tilde
G\) +(2\eta-1) \(G'G + \tilde G'\tilde G\) = 0~, &\eqsGGtilde a
\cr &(\eta-1)\(G' G' - G'' G - \tilde G'\tilde G'+\tilde G''\tilde
G\) - \(G'G - \tilde G'\tilde G\) = 0~, &\eqsGGtilde b \cr
&\eta\(\tilde G''G + \tilde G G'' - 2\tilde G' G'\) + \(\tilde G'G
+ \tilde G G'\) = {\mR^2-1/4\over(1-\eta)^2}\,\tilde G G~,
&\eqsGGtilde c
}$$
where primes denote derivatives with respect to $\eta$ and the first two
equations \eqsGGtilde {a,b} are consequences of Eq.~\eqI.

It is easy to see that for the auxiliary functions $\chi(\eta)$,
$\varphi(\eta)$ introduced in \param, Eqs.~\eqsGGtilde {} become
\eqna\phichieq$$\eqalignno{
&\varphi''\(1-\eta\cosh^2\varphi\)
+\h\varphi'^2\eta\sinh2\varphi
+ \varphi'\,{1-2\eta+\eta^2\cosh^2\varphi\over\eta(1-\eta)} =
{\mR^2-1/4\over2\eta(1-\eta)}\sinh2\varphi ~, \indent&\phichieq a
\cr &\chi'\(1-\eta\cosh^2\varphi\) = \varphi'^2\,\eta(1-\eta) +
\h\varphi'\,\eta\sinh2\varphi
-{\mR^2-1/4\over1-\eta}\sinh^2\varphi~, &\phichieq b
}$$
which immediately imply the representation in terms of the Painlev\'e VI
transcendent in \painleve, \phichieqintro. In the flat-space limit
$R\to\infty$ these equations reduce to the well-known sinh-Gordon form.

The short distance asymptotic behaviour $\eta\to 0$ of the appropriate
solution can be obtained by iteratively solving Eqs.~\phichieq{}
with prescribed leading behaviour \wsmall, \ordernorm. In terms of the
combinations $F_-(\eta) = (2R)^{1/4}\eta^{1/8}\,G(\eta)$ and
$F_+(\eta) = (2R)^{1/4}\eta^{1/8}\,\tilde G(\eta)$, we obtain
\eqn\short{\eqalign{
F_\pm(\eta)
= 1 &\pm\mR\,\eta^{1/2}\Omega +
{1\over8}(2\mR^2-1)\eta
\pm{1\over8}\mR\,\eta^{3/2}\[2+\Omega+2\mR^2\Omega\]
\cr
&+{1\over128}\eta^2
\[-7+\mR^2\(15+8\Omega^2\) + \mR^4\(1+8\Omega-8\Omega^2\)\]
\cr
&\pm{1\over128}r\,\eta^{5/2}
\[17+7\Omega+3r^2(3+7\Omega)+r^4(-1+5\Omega)\]
+O\(\eta^3\,\Omega^2\)~,
}}
where the term $\Omega$ contains the logarithmic dependence in $\eta$,
\eqn\omegadef{
\Omega = \ln \(k\,\eta^{1/2}\)~.
}
The coefficient $k$ above, given by Eq.~\logk, is related to the
expectation value of the energy field \energyvev\ as described in
Appendix~A.

Alternatively, this solution can be specified by the large distance
asymptotic behaviour $\eta\to 1$ with leading behaviour \wlarge\ and
$\chi(\eta)\sim4\ln\bar s$. Eqs.~\phichieq{} then yield
\eqn\GGtildelarge{\eqalign{
&{G(\eta)\over\bra\sigma\ket^2}
= 1 + A^2(1-\eta)^{1+2\mR} g_2(1-\eta)
+ A^4(1-\eta)^{2+4\mR} g_4(1-\eta)
+ O((1-\eta)^{18+6r}),
\cr
&{\tilde G(\eta)\over\bra\sigma\ket^2}
= A(1-\eta)^{1/2+\mR} \tilde g_1(1-\eta)
+ A^3(1-\eta)^{3/2+3\mR} \tilde g_3(1-\eta)
\cr&\hskip137pt
+ A^5(1-\eta)^{5/2+5\mR} \tilde g_5(1-\eta)
+ O((1-\eta)^{49/2+7r}),
}}
where the constant $A$, given in \coefflarge, can be obtained from the
one-particle form factors; the magnetization $\bra\sigma\ket$ is given in
\magnetization;
\eqn\Glargeg{\eqalign{
g_2(x) &= {1+2\mR\over2^6(1+\mR)}
\[{x\over1+\mR} + x^2 +
{(3+2r)(39+34r+8r^2)\over32(2+r)^2}x^3 + O(x^4)
\]~,
\cr g_4(x) &=
{3(1+2\mR)^3(3+2\mR)\over2^{24}(1+\mR)^4(2+\mR)^3(3+\mR)^2}
\bigg[{x^6\over2+\mR} + 2x^7
+O(x^8)\bigg]~;
}}
and
\eqn\Gtildelargeg{\eqalign{
\tilde g_1(x) =& {1\over2}F\(\th+\mR,\th+\mR;1+2\mR;x\)~,
\cr
\tilde g_3(x) =& {(1+2\mR)^2(3+2r)\over2^{13}(1+\mR)^3(2+\mR)^2}
\left[{x^3\over3+2r}+{3\over4}x^4
+{3(221+255r+96r^2+12r^3)\over64(3+r)^2}x^5
+O(x^6) \right]~,
\cr
\tilde g_5(x) =& {9(1+2\mR)^4(3+2\mR)^2(5+2r)
\over2^{36}(1+\mR)^5(2+\mR)^5(3+\mR)^4(4+\mR)^2}
\bigg[{x^{10}\over5+2r}+{5\over4}x^{11}
+ O(x^{12})\bigg]~.
}}
The functions $g_6(1-\eta)$ and $\tilde g_7(1-\eta)$ that enter in
the terms $A^6(1-\eta)^{3+6r}$ and \hbox{$A^7(1-\eta)^{7/2+7r}$}
of $G$ and $\tilde G$ are of order $(1-\eta)^{15}$ and
$(1-\eta)^{21}$, respectively. These powers are simple
consequences of the Pauli exclusion principle on the particle
states of the Hilbert space, as will be apparent in the next
sub-sections, and are just the sums of the first integers up to 5
and up to 6, respectively.

It is possible to check order by order in both of these expansions that
this solution satisfies the ``duality'' relation
\eqn\GGtildedual{
\bra\sigma(x)\sigma(x')\ket\Big|_{r=\pm\h}
= \bra\mu(x)\mu(x')\ket\Big|_{r=\mp\h}~,
}
as depicted in Figure~2. The relation \GGtildedual\ should not come as a
suprise given that,  as described in the Introduction, for $r<-\h$ we
indeed have to trade
$\sigma\leftrightarrow\mu$ due to the fact that only the ``free''
asymptotic condition $\bra\sigma\ket=0$ is stable. Nevertheless, while
\GGtildedual\ is verified by the short
distance expansion \short\ essentialy because of the property $k(1/2) =
k(-1/2)$, for $k(r)$ in \logk, its validity from the large distance
expansion \GGtildelarge\ point of view is far less trivial.

Asymptotics for the auxiliary functions $\chi(\eta)$, $\varphi(\eta)$
can be found in Appendix~D.

\nobreak
\midinsert
\centerline{\vbox{
\hbox {\hsize 16cm
\hskip-0.5cm
\vbox{\hsize 8cm
\centerline{\epsfysize=4.5cm\epsfbox{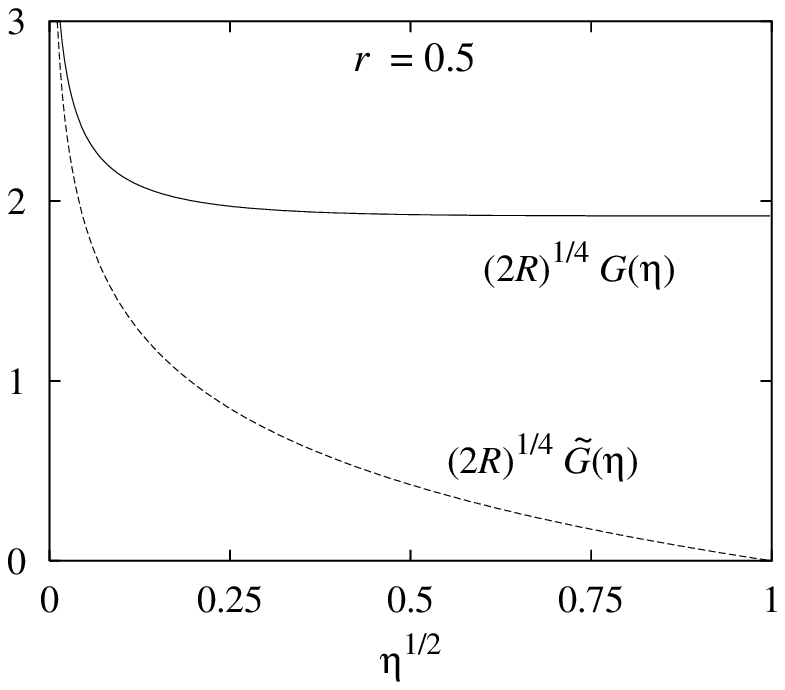}}
\vskip-3pt
\centerline{\hskip0pt{\nineb({\nineib a})}}
}
\hskip-1cm
\vbox{\hsize 8cm
\centerline{\epsfysize=4.5cm\epsfbox{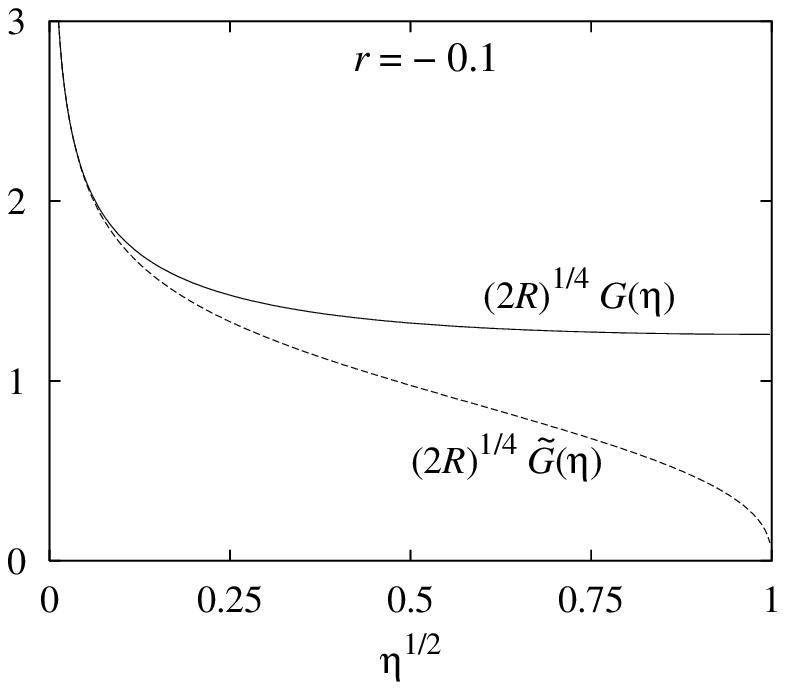}}
\vskip-3pt
\centerline{\hskip0pt{\nineb({\nineib b})}}
}}
\bigskip
\hbox {\hsize 16cm
\hskip-0.5cm
\vbox{\hsize 8cm
\centerline{\epsfysize=4.5cm\epsfbox{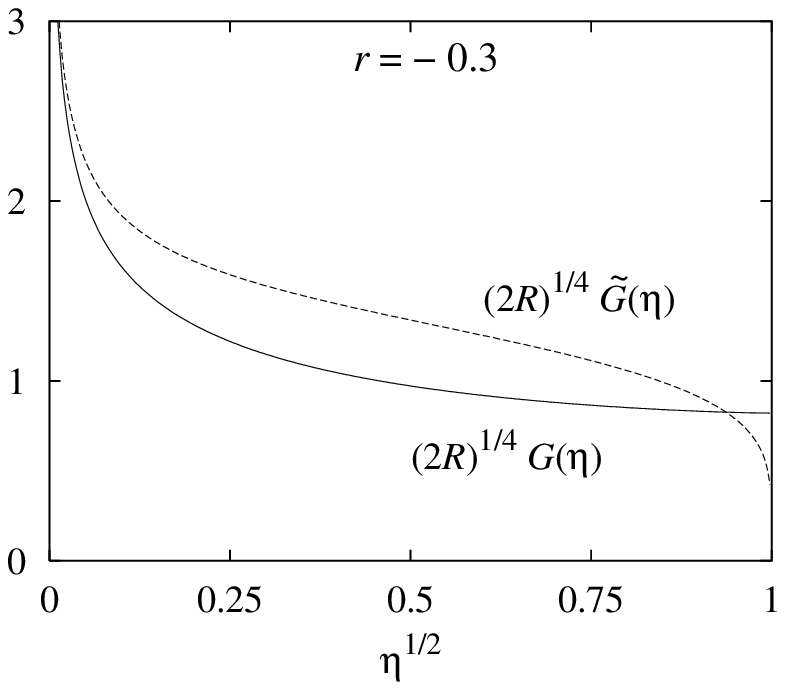}}
\vskip-3pt
\centerline{\hskip0pt{\nineb({\nineib c})}}
}
\hskip-1cm
\vbox{\hsize 8cm
\centerline{\epsfysize=4.5cm\epsfbox{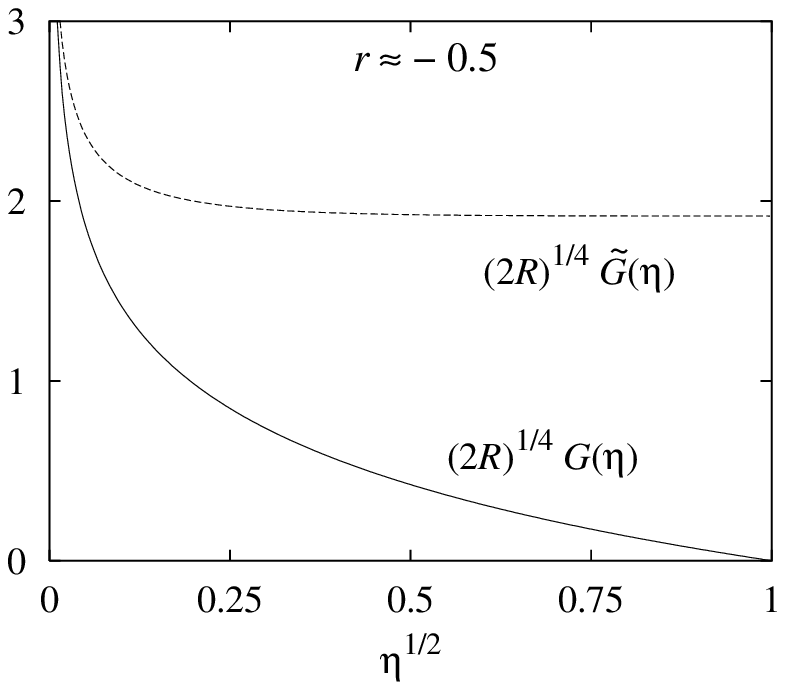}}
\vskip-3pt
\centerline{\hskip0pt{\nineb({\nineib d})}}
}}
\begingroup\parindent=0pt\leftskip=1cm\rightskip=1cm\parindent=0pt
\baselineskip=12pt \ninerm \ninemath
\vskip10pt
\vskip 6pt
{\nineb Fig.~2:}~Plots of the spin-spin correlation functions $G(\eta)$,
$\tilde G(\eta)$, Eqs.~\GGtilde, against the distance $\eta^{1/2}$ for
some
values of
the parameter $\mR$. These were obtained from numerical integration of
Eqs.~\eqsGGtilde {} using the short distance expansion \short;
agreement was found when comparing to the large distance behaviour
\GGtildelarge; ({\nineib a}) $r = 0.5$; ({\nineib b})~$r= -0.1$;
({\nineib c})~$r= -0.3$; ({\nineib d})~$r\approx -0.5$.
\vskip 6pt
\endgroup
}}
\endinsert
\goodbreak

\subsec{The Form Factors}

For our purpose it is enough to consider form factors of fields
placed at the center of the Poincar\'e disk. And because we are
considering the ``ordered'' case $\bra\sigma\ket\neq0$, form factors of
the form $\bra{\rm vac}|\sigma(0)|n_1\ldots n_{2N+1}\ket$, $\bra{\rm
vac}|\mu(0)|n_1\ldots n_{2N}\ket$ trivially vanish.

One-particle form factors can be obtained from the
Ward identity associated with the invariance of a single Ising system under
rotations around the origin,
\eqn\wardone{
\bra{\rm vac}|\, \big[{\bf R},\,\mu(0)\big]\,|n\ket = 0~.
}
Using the action of ${\bf R}$ on particle states, Eq.~\Rstate, this yields
the recursion relation
\eqn\eqone{
\bra{\rm vac}|\mu(0)|n+2\ket =
{\alpha_n\over\alpha_{n+1}}\bra{\rm vac}|\mu(0)|n\ket~,
}
with coefficients $\alpha_n$ given in \alphan. Hence, $\bra {\rm
vac}|\mu(0)|2n+1\ket = 0 $ and
\eqn\oneeven{
\bra{\rm vac}|\mu(0)|2n\ket =
\bra{\rm vac}|\mu(0)|0\ket
\(\Gamma\(n+\th\)\(\mR+\th\)_n\over
\sqrt\pi\,\Gamma(n+1)(\mR+1)_n\)^{1/2} ~,
}
where $n=0,1,2,\ldots$ and $(a)_n = \Gamma(a+n)/\Gamma(a)$. The
one-particle form factor \hbox{$\bra{\rm vac}|\mu(0)|n=0\ket$} is
determined from the short distance behaviour of
\eqn\psimushort{
\bra \psi(x)\,\mu(0)\ket \sim \bra\sigma\ket\,\sqrt{-i\over2 z}
\indent\hbox{as} \indent |x|\to 0~.
}
Using the expansion for $\psi(x)$ in \waves\ and re-summing leading large
$n$ behaviour of \oneeven, we find
\eqn\onezero{
\bra{\rm vac}|\mu(0)|0\ket = \bra\sigma\ket
\(\Gamma\(\mR+\th\)\over\sqrt\pi\,\Gamma(\mR+1)\)^{1/2}~,
}
which agrees with the result obtained in Ref.~\DoyonII.

Two-particle form factors can be obtained from the Ward identity for
the doubled system,
\eqn\wardtwo{
\bra{\rm vac}|\,\big[{\bf Y}_1 - {\bf Y}_{-1},
\,\mu_a(0)\mu_b(0)\big]\,|n_1\,n_2\,;\,\vac\ket
= i\bra{\rm vac}|\,\big[{\bf H}_a - {\bf H}_b,
\,\sigma_a(0)\sigma_b(0)\big]\,|n_1\,n_2\,;\,\vac\ket\,,
}
coming from \Yaction. The combination ${\bf Y}_1 - {\bf Y}_{-1}$ can be
expressed in terms of ${\bf X}_1 - {\bf X}_{-1}$ and ${\bf Z}_0$ as in
\chargesY, so that it is straightforward to use the results from
Section 2 to evaluate its action on the particle state. Using
\eqone\ as well, the equation above can be written as
\eqn\eqtwo{\eqalign{
-i\(\omega_{n_1} + \omega_{n_2}\)
\bra\sigma\ket\bra\vac|\sigma(0)|n_1\, n_2\ket
&=\alpha_{n_2}\bra\vac|\mu(0)|n_1\ket\bra\vac|\mu(0)|n_2+1\ket
\cr
&-\alpha_{n_1}\bra\vac|\mu(0)|n_2\ket\bra\vac|\mu(0)|n_1+1\ket~,
}}
and hence the non-zero elements are
\eqn\twoevenodd{
\bra\sigma\ket\bra\vac|\sigma(0)|2n_1,2n_2+1\ket =
{i\,\alpha_{2n_2}\over \omega_{2n_1}
+ \omega_{2n_2+1}}\bra\vac|\mu(0)|2n_1\ket\bra\vac|\mu(0)|2n_2\ket~,
}
where $\omega_n$ and $\alpha_n$ are given, respectively, by
\Heigenvalue\ and \alphan. This formula agrees with the results found in
Ref.~\DoyonII.

Form factors involving three or more particles can be written in
terms of the one-particle and two-particle form factors using
Wick's theorem, which can in fact be seen as a consequence of Ward
identities as above. For instance, in the case of the three- and
four-particle form factors, we can use Eqs.~\Qaction\ and the
action of the $U(1)$ charge ${\bf Z}_0$ on particle states
\actionQstates. The Ward identities
\eqn\wardwick{\eqalign{
&\bra{\rm vac}|\, \big[{\bf
Z}_0,\,\sigma_a(0)\mu_b(0)\big]\,|n_1\, n_2\, n_3\,;\,\vac\ket
=-i\bra\vac|\mu_a(0)\sigma_b(0)|n_1\, n_2\, n_3\,;\,\vac\ket~, \cr
&\bra{\rm vac}|\,\big[{\bf Z}_0,
\,\sigma_a(0)\sigma_b(0)\big]\,|n_1\, n_2\, n_3\,;\,n_4\ket = 0
}}
give, respectively,
\eqn\eqwick{\eqalign{
\bra\sigma\ket \bra\vac|\mu(0)|n_1\, n_2\, n_3\ket
&=
  \bra\vac|\mu(0)|n_1\ket\,\bra\vac|\sigma(0)|n_2\,n_3\ket
+ \bra\vac|\mu(0)|n_3\ket\,\bra\vac|\sigma(0)|n_1\,n_2\ket \cr &\,
+ \bra\vac|\mu(0)|n_2\ket\,\bra\vac|\sigma(0)|n_3\,n_1\ket
}}
and
\eqn\fourparticles{\eqalign{
\bra\sigma\ket\bra\vac|\sigma(0)|n_1\, n_2\, n_3\, n_4\ket &=
\bra\vac|\sigma(0)|n_1\, n_2\ket\bra\vac|\sigma(0)|n_3\, n_4\ket
\cr
&-\bra\vac|\sigma(0)|n_1\, n_3\ket\bra\vac|\sigma(0)|n_2\, n_4\ket
\cr
&+\bra\vac|\sigma(0)|n_1\, n_4\ket\bra\vac|\sigma(0)|n_2\, n_3\ket~.
}}

The large distance expansion for the two-point correlation functions
\GGtilde\ can be obtained by inserting a complete set of states between
the two operators and by expressing them as summations over form factors
contributions,
\eqn\GGtildelargeII{\eqalign{
G(\eta) &= \bra\sigma\ket^2 + \sum_{n_1,n_2 \ge 0}
|\bra\vac|\sigma(0)|2n_1,2n_2+1\ket|^2~
e^{-(\omega_{2n_1}+\omega_{2n_2+1}){\dist \over2R}}
\cr
&\hskip32pt
+~~ \hbox{$(2l)$--particle contributions }(l\ge2)~,
\cr
\tilde G(\eta) &= \sum_{n\ge0} |\bra\vac|\mu(0)|2n\ket|^2~
e^{-\omega_{2n}{\dist \over 2R}}
~+~~ \hbox{$(2l+1)$--particle contributions }(l\ge1)~,
}}
where $\dist$ denotes the geodesic distance, related to $\eta$ by
\projinvdist, and $\omega_n$ is the energy eigenvalue
\Heigenvalue. As noted in \DoyonI, these expressions are obtained
using $SU(1,1)$ transformations in order to bring one of the spin
operators to the origin of the disk, the other to the imaginary
axis $z=iq,\,\bar z=-iq$, and by using the representation
\eqn\heisenberg{
\sigma(iq,-iq) = e^{{\bf H} {\dist\over2R}}\,\,\sigma(0)\,\,e^{-{\bf H}
{\dist\over2R}}~,
}
where ${\bf H}$ is the Hamiltonian \Hcharge\ and $\dist$ is the
geodesic distance from the point $(iq,-iq)$ to the origin. The
expansions \GGtildelargeII\ agree with the asymptotics
\GGtildelarge\ obtained from the analysis of the differential
equations, with the coefficient $A$ being related to the
one-particle form factor $\bra\vac|\mu(0)|0\ket$ in \onezero\ by
\eqn\coefflargeII{
A\,\bra\sigma\ket^2 = 4^{-\mR}\bra\vac|\mu(0)|0\ket^2~,
}
which gives \coefflarge.

\newsec{Thermodynamics and Discussion}

It is natural to assume that the Ising quantum field theory on the
pseudosphere represents the scaling limit of an Ising-like
statistical system on a lattice embedded into the pseudoshpere.
Although we do not have yet a precise construction of this
statistical system and of its scaling limit, it is interesting to
interpret our results assuming that most of the basic concepts
underlying the scaling limit in the flat-space situation carry
over to the pseudosphere. In particular, one can assume that in
the process of the scaling limit, the curvature of the
pseudosphere is brought towards zero in microscopic units, and
that the lattice embedded into the pseudosphere resembles more and
more, in regions of radius much smaller than $R$, a prescribed
(regular) flat-space lattice. At the same time, the temperature is
brought towards the critical temperature of this prescribed
flat-space lattice, with a prescribed ratio between $R$ and the
resulting flat-space correlation length. This ratio is given by
the parameter $\mR = mR$ that we introduced for the Ising quantum
field theory on the pseudosphere.

\subsec{The Free Energy}

An interesting quantity to study is the specific free energy
$f(m,R)$ as function of the mass $m$ and curvature radius $R$,
defined through the partition function $Z$ by \hbox{$f =
-\lim_{V\to\infty} \ln Z^{1/V}$} where $V$ is the
(two-dimensional) volume of space.

A particularly simple case is the massless one, $m=0$, where the
free energy $f(m=0,R)$ can be computed using the defining relation
between the trace of the energy-momentum tensor $T_\mu^\mu$ and
the variation of the action $S$ under a scale transformation of
the metric $g_{\mu\nu}$, $\sqrt{g}\, T^\mu_\mu = 2 g_{\mu \nu}\,
\delta S /\delta g_{\mu\nu}$. This gives
\eqn\anomaly{
R\frc{d}{dR} \Big[V(R)\,f(0,R)\Big] = V(R)\<T^\mu_\mu\>~,
}
where the volume $V(R)$ must be taken finite (but large) for this
equation to make sense and must vary like $R^2$ under a scale
transformation, $R\, dV(R)/dR = 2V(R)$. For the pseudosphere, the
trace anomaly is related to the central charge $c$ by
$\<T^\mu_\mu\> = c/(12 \pi R^2)$, where we have set to zero the
constant piece related to the vacuum energy density. With $c=1/2$,
this yields
\eqn\freeenergyconformal{
    f(m=0,R) = \frc{1}{24\pi R^2} \ln\(\frc{2R}{L}\)~,
}
where $L$ is an integration constant not determined by the quantum
field theory but only fixed after specifying the microscopic
theory, i.e. the theory with an explicit ultra-violet cutoff. In
the scaling limit of the corresponding microscopic theory, that
is, setting the temperature equal to the flat-space critical
temperature and making $R$ very large in microscopic units, the
corresponding free energy per unit volume is expected to have the
leading behaviour \freeenergyconformal.

Such a geometrical contribution to the specific free energy, and
in particular the presence of the non-universal distance $L$, is
expected for theories on a space that is not asymptotically flat.
In the case of the pseudosphere, the logarithmic increase of
$R^2f(0,R)$ as $R$ increases is related to the decrease of the
``space available'' around every site, which decreases the
interaction energy and increases the free energy (as opposed to
the case of a sphere of radius $R$, where one observes a decrease
of $R^2 f(0,R)$ \AlZamo). In comparison, there is no such
contributions to the specific free energy in asymptotically flat
spaces without singularities, for instance. There is only a finite
and universal contribution to the total, volume-integrated free
energy; for a conformal-to-flat metric $g_{\mu\nu} = e^{2\phi}
\delta_{\mu,\nu}$, this contribution is given by the well-known
formula
\eqn\weelknownformula{
    -\frc{c}{24\pi} \lim_{V\to\infty} \int_V d^2x\,
    \d_\mu\phi\, \d^\mu\phi~.
}

The free energy $f(m,R)$ of the Ising field theory at arbitrary
mass can be obtained from the vacuum expectation value of the
energy field at finite volume $V$, here denoted by
$\bra\varepsilon\ket_V(x)$, by taking the infinite-volume limit:
\eqn\freeenergylimit{
    \frc{d}{d m} f(m,R)
= -\frc1{2\pi} \lim_{V\to\infty} \frc1V \int_V d^2x
    \lt[ \bra\varepsilon\ket_V(x) + m \ln\(\frc{2R}\epsilon\) \rt]~,
}
where $\epsilon$ is another non-universal microscopic distance.

In \freeenergylimit, it is tempting to take the limit of infinite
volume $V$ by simply replacing $\bra\varepsilon\ket_V(x)$ by its
infinite-volume, position-independent expression \energyvev.
However, because on the pseudoshpere the surface enclosing a
finite region increases as fast as its volume for large volumes,
it is possible that contributions proportional to the surface,
arising from integration of $\bra\varepsilon\ket_V(x)$ at
positions $x$ near the boundary (where it is significantly
different from $\energyvev$), give in $f(m,R)$ additional finite
terms. We have not yet evaluated these contributions, but
expect to come back to this problem in a future work.

A similar situation was found in the study of the Ising model on
hyperlattices \Rietman. As the authors did, we focus our attention
on a ``bulk'' free energy, defined by taking for
$\bra\varepsilon\ket_V(x)$ in \freeenergylimit\ the expression
\energyvev, valid at positions $x$ far from the boundary. This
gives
\eqn\freeenergy{
    2\pi R^2 f(m,R) = \ln G(1+\mR) - \frc\mR2 \ln2\pi
    - \frc{\mR^2}2 \ln\({ 2R \over e^{1+\gamma}
    \epsilon}\) +
    \frc1{12} \ln\(\frc{2R}{L}\)~,
}
where $G(z)$ is Barnes' $G$-function
\eqn\barnes{
    G(z+1) = (2\pi)^{z/2} e^{-[z(z+1)+\gamma z^2]/2}
    \prod_{n=1}^\infty \lt[ \(1+\frc{z}n\)^n e^{-z+z^2/(2n)} \rt]~,
}
and $\gamma$ is Euler's constant. This expression has the small~$\mR$
convergent expansion
\eqn\freeenergysmall{
    2\pi R^2 f(m,R) = \frc1{12} \ln\(\frc{2R}L\) - \frc{\mR}2 -
    \frc{\mR^2}2 \ln\(\frc{2R}{\epsilon}\) + \sum_{n=3}^\infty
    (-1)^{n-1} \zeta(n-1) \frc{\mR^n}n~,
}
where $\zeta(n)$ is Riemann's zeta function, as well as the following
asymptotic expansion at large $\mR$,
\eqn\freeenergylarge{
    2\pi R^2 f(m,R)  = \frc{\mR^2}2 \ln\( \frc{m\,\epsilon\,
    e^{\gamma-1/2}}2\) - \frc1{12}
    \ln \(\frc{m\,L}2\) + \zeta'(-1) +
    O\(\mR^{-2}\)~,
}
with the first term corresponding to the specific free energy of
the massive Majorana fermion theory in flat space (the
$R\to\infty$ limit). As on the sphere \AlZamo, there is no
logarithmic term in $R$ in this large $\mR$ expansion.

From the analytic properties of Barnes' $G$-function, one can see
that the free energy \freeenergy\ has logarithmic singularities
located at the negative integers $\mR=-1,\,-2,\,\ldots$. In
particular, it is regular at $\mR=0$, that is, the flat space
critical temperature does not correspond to a singularity of the
free energy.

It is interesting to note that when we fix $\mR = 1/\sqrt6$,
the free energy \freeenergy\ does not depend on $R$ anymore, and
only the ratio $L/\epsilon$ appears. In this case, the logarithmic
increase of $R^2 f(m,R)$ as $R$ increases due to effects of the
geometry as explained above is exactly cancelled out by the
logarithmic decrease due to the increase of the interaction energy
as the correlation length grows.

The ``bulk'' free energy defined above still depends on the
asymptotic conditions of the quantum field theory. Specifically,
the expression \freeenergy\ is valid for ``fixed'' asymptotic
conditions, whilst the replacement $\mR\mapsto -\mR$ gives the
expression for ``free'' asymptotic conditions. Both asymptotic
conditions, or regimes, are stable in the region $-\h<\mR<\h$ and
we intend to discuss the possible transitions between these
regimes in a future work. A full treatment of the thermodynamics
of the model, in fact, seems to require a better understanding of
the nature and importance of the neglected surface terms as well
as of the other stable asymptotic conditions that break part or
all of the symmetries associated to the isometry of the
pseudosphere, as described in Section~2.2.

\subsec{The Magnetization}

The expression \magnetization\ for the magnetization
$\bra\sigma\ket$ in the Ising field theory is expected to
determine the coefficient of the leading asymptotic behaviour of
the magnetization in the microscopic theory as the scaling limit
is taken. As depicted in Figure~3, it does not vanish at the flat
space critical temperature $m=0$, but rather at a higher
temperature, corresponding to the value $m=-\frc1{2R}$ of the mass
parameter. That is, at $\mR = -1/2$, there is a change in the
power law of the leading asymptotic behaviour of the magnetization
in the microscopic theory as the scaling limit is taken. From this
only, we cannot conclude that there exists an $R$-dependent
temperature at which the magnetization vanishes identically in the
microscopic theory for any finite $R$. However, the vanishing of
the magnetization occurs at the value of $m$ below which the
ordered regime is unstable and the disordered regime, where the
magnetization is zero, is stable. It is plausible that there be a
similar point joining an ordered and a disordered regime at finite
$R$ in the microscopic theory at a temperature higher than the
flat-space critical temperature (higher by an amount that has the
power law behaviour $\sim R^{-1}$ as the scaling limit is taken).
The magnetization would vanish at the turning point between the
two regimes, as it has been suggested for the regular lattice
theory studied in Ref.~\Rietman. We note, though, that our
expression \freeenergy\ for the free energy is regular at $\mR =
-1/2$. Of course, as we have pointed out, the expression
\freeenergy\ probably does not give the full free energy, hence no
serious conclusion can be drawn from it yet.

\fig{3}{Plot of the magnetization $(2R)^{1/4}\bra\sigma\ket^2$, in
Eq.~\magnetization. The filled line corresponds to the choice of ``fixed''
boundary condition for the Ising field theory \action\ in the region
$r>-\h$, whereas the dotted line corresponds to the choice of ``free''
boundary condition for $r<\h$.}
{\epsfysize=4.5cm\epsfbox{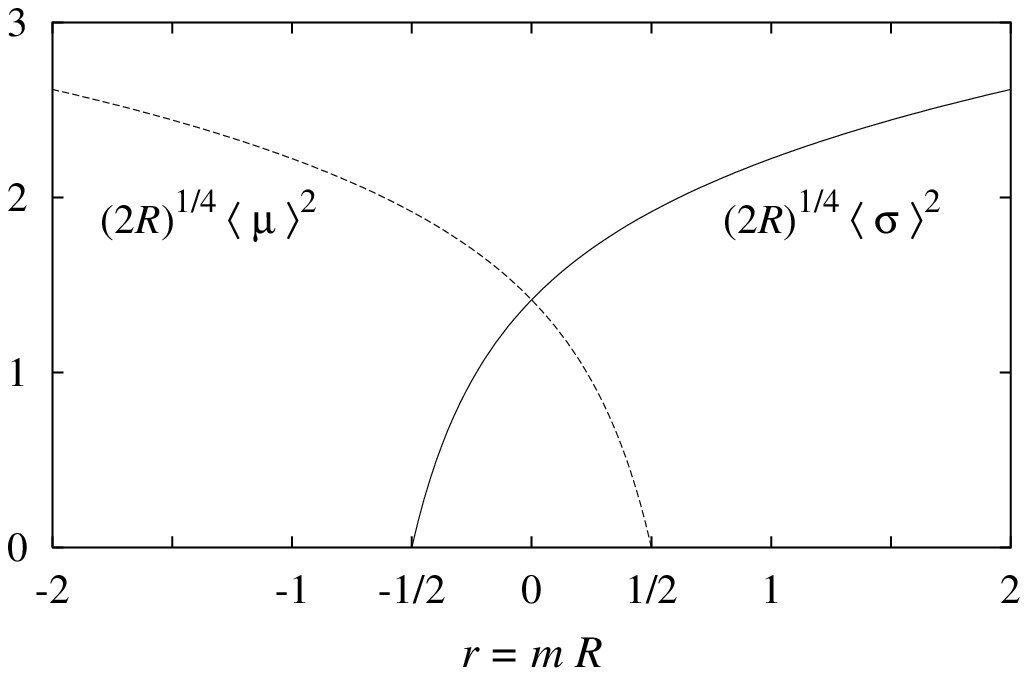}}

Near the effective ``critical'' temperature, the magnetization in
the Ising field theory vanishes as
\eqn\dualpoint{
\bra\sigma\ket^2 = (2R)^{-1/4}\,
\sqrt\pi\,\bar s_{\rm flat}^4\,\(\mR+\th\) + O\(\(\mR+\th\)^2\)~,
}
where $\bar s_{\rm flat} = 2^{1/12}\,e^{-1/8}\,A^{3/2}$ ($A$ being
Glaisher's constant). The exponent $1/2$ can be explained by
recalling that a space of constant negative curvature is
effectively infinite-dimensional at large distances due to the
fact that the volume grows exponentially \Callan. In fact, a
theory on the pseudosphere should essentially show, in some sense,
a cross-over behavior from a two-dimensional theory to an
infinite-dimensional theory. Hence mean-field theory could be used
to predict the exponent ruling the vanishing of the magnetization
in the Ising field theory, giving $1/2$ as above. Assuming that
the magnetization in the microscopic theory vanishes similarly at
a ``critical'' temperature, the exponent ruling its vanishing
should then be $1/2$, which agrees with the results of
Ref.~\Rietman. In the flat-space limit $R\ra\infty$, the
magnetization takes the usual form $\bra\sigma\ket \ra \bar s_{\rm
flat}\,m^{1/8}$, and the exponent ${1/8}$ is recovered.

The fact that the effective ``critical'' temperature is higher
than the flat-space critical temperature is expected: the
asymptotic conditions have a greater effect on the pseudosphere
than they have on flat space, hence the ``fixed'' asymptotic
conditions will render the establishment of disorder more
difficult, increasing the effective ``critical'' temperature.
Similar considerations apply in the disordered regime: there the
effective temperature at which the average of the disorder
variable vanishes is lower than the flat-space critical
temperature, because ``free'' asymptotic conditions make it more
difficult to establish order.

\subsec{Two-point Correlation Functions and Susceptibility}

An interesting characteristics of the two-point functions is their
exponential decay at large geodesic distances,
\eqn\Gexpdecay{
    \frc{\bra\sigma(x)\sigma(x')\ket}{\bra\sigma\ket^2} - 1 \sim
    \frc{\Gamma\(\frc12+\mR\) \Gamma\( \frc32 +
    \mR\)}{2\pi\,\Gamma^2\(2+\mR\)} \; e^{-2(1+\mR)
    \,d(x,x')/R} \quad\hbox{as}\quad \dist(x,x')\ra\infty~,
}
and
\eqn\Gtexpdecay{
    \frc{\bra\mu(x)\mu(x')\ket}{\bra\sigma\ket^2} \sim
    \frc{\Gamma\(\frc12+\mR\)}{\sqrt\pi\,\Gamma\(1+\mR\)}\;e^{-\(\h+\mR\)
    \,d(x,x')/R} \quad\hbox{as}\quad \dist(x,x')\ra\infty~.
}
Contrary to the flat-space case, the exponential decay is
different for the order-order and for the disorder-disorder
two-point functions. Hence one cannot define a unique correlation
length valid for describing the long distance behaviour of both
correlation functions. It is natural, however, to choose the
exponential decay of the disorder-disorder two-point function as
the one defining an effective correlation length,
\eqn\corrl{
    \xi = \frc{2R}{1+2\mR}~.
}
We indeed expect this correlation length to diverge at the point
$\mR = -\h$ in the ordered regime where the magnetization
$\bra\sigma\ket$ vanishes, since at this point, the disorder field
acquires a nonzero expectation value and the large distance
asymptotic behaviour of its two-point function changes. This
correlation length diverges as the inverse power of the difference
of the temperature to the effective ``critical'' temperature, as
is the case for the Ising model on flat space, but it is defined
here only for the behaviour from above the point $\mR = -\h$,
since below this point the system is necessarily in its disordered
regime. A corresponding definition of the correlation length in
the disordered regime leads to a divergence at the point $\mR =
\h$ from below. Following considerations similar to those of the
previous subsection, we expect to have the same power law
behaviour of the correlation length in the lattice theory in the
vicinity of the critical point. Note that a naive application of
general results from mean field theory would predict the power law
$\sim (1+2\mR)^{-1/2}$.

In the ordered regime ($\mR>-\h$), and as the point $\mR = -\h$ is
approached, the two-point function of disorder fields goes to an
almost constant value before vanishing at larger and larger
distances. This almost constant value approaches the value that
$\bra\mu\ket^2$ takes in the disordered regime at $\mR = -\h$.
More precisely, as $\mR=-\h$ is approached in the ordered regime,
both order-order and disorder-disorder two-point functions tend to
the exact form they have in the disordered regime at $\mR = -\h$
(see Figure 2). This is a consequence of the duality relating the
point $\mR =-\h$ to the point $\mR = \h$, which yields
Eq.~\GGtildedual. A similar duality has also been observed in the
study of the statistical Ising model on a hyperlattice~\Rietman.

It is also interesting to consider the general case where an
external magnetic field $h$ is added to the Ising field theory
\action, by adding the perturbation $h\int
d^2x~e^{\phi(x)}\,\sigma(x)$. The corresponding susceptibility
$\chi$ giving the linear response of the magnetization is given by
\eqn\susc{\eqalign{
    \chi &= \h \int \frc{d^2x}{4R^2} ~ e^{\phi(x)}\,
    \(\bra\sigma(x)\sigma(0)\ket - \bra\sigma\ket^2\)
\cr
    &= \frc{\pi}{4R} \int_0^\infty ds ~ \sinh\(\frc{s}R\) \,
    \(\bra\sigma(x)\sigma(0)\ket\big|_{\dist(x,0)=s} -
    \bra\sigma\ket^2\)~.
}}
Using the asymptotic behaviour \Gexpdecay\ for the ordered regime,
it is straightforward to see that the integral above is convergent
for any $\mR>-\h$, with a divergence at $\mR=-\h$,
\eqn\suscdiv{
    \chi \sim \frc{\bar s_{\rm flat}^4}{4\sqrt\pi}
\,{(2R)^{-1/4}\over1+2\mR}
\indent\hbox{as}\indent\mR\ra-\h~.
}
The susceptibility again shows a divergence at the effective
``critical'' value $m=-\frc1{2R}$, with a mean-field power law
behaviour. A similar phenomenon was observed for the model studied
in \Rietman. In the disordered regime, using the asymptotic
behaviour \Gtexpdecay\ with $\mR \mapsto -\mR$ for the order-order
two-point function, one can see that the susceptibility is finite
for $\mR< -\h$ and diverges with mean-field power law at
$\mR=-\h$, even though the regime is stable above $-\h$. Hence in
the whole range $-\h<\mR<\h$ in the disordered regime, the
response of the magnetization to a magnetic field is not linear at
small magnetic field. Note also that the susceptibility diverges
at $\mR=-\h$ from both directions with the same exponent.

Relating the susceptibility to the expansion of the free energy in
powers of the magnetic field in the usual fashion, one could
conclude from this analysis that the free energy possesses a
singular behaviour at small magnetic field in the region
$-\h<\mR<\h$ of the disordered regime. However, from
considerations similar to those above, it is possible that one
needs to take into account surface terms in order to obtain the
correct coefficients in the expansion of the free energy in powers
of magnetic field. We hope to carry out this analysis in a future
work.

\vskip 0.3in

\centerline{\bf Acknowledgments}

We are grateful to S. Lukyanov and A. Zamolodchikov for helpful
discussions. B.D. also acknowledges insightful comments from J.
Cardy, and P.F. is thankful for useful discussions with Y.~Saint-Aubin and
Al.~Zamolodchikov as well as for the conversations with R.~Conte,
A.~Fateev, I.~Kostov, V.~Pasquier, J.~Rasmussen, H.~Saleur and
V.~Schomerus.  P.F. is also extremely grateful to the New High Energy 
Theory Center at Rutgers University where a great part of this work was 
done.

The work of P.F. was partially supported by the grants
DOE \#DE-FG02-96 ER 40959 and POCTI/FNU/38004/2001-FEDER (FCT, Portugal).

\vskip 0.1in

\appendix{A}{}

The vacuum expectation value of the energy field $\varepsilon(x) =
i(2R)^{-1}(1-z\bar z)(\psi\bar\psi)(x)$ can be obtained from the
propagator
\eqn\propagator{
\bra\,\psi(x)\,\bar\psi(x')\,\ket = (1-z\bar z')^{-1} {\cal G}(\eta)~,
}
where the piece
\eqn\calG{
{\cal G}(\eta) =
{\Gamma(\mR)\Gamma(1+\mR)\over2i\,\Gamma(2\mR)}(1-\eta)^\mR
F(\mR,1+\mR;1+2\mR;1-\eta)~,
}
is function of the projective invariant $\eta$, Eq.~\projinv. This
propagator is determined by the equations of motion \eqmotion\
\eqn\eqpropagator{
(1-z\bar z)\;\partial_z \Big( (1-z\bar z) \partial_{\bar z}
\bra\psi(x)\bar\psi(x')\ket\Big)
= \mR^2 \bra\psi(x)\bar\psi(x')\ket~,
}
and the normalization condition
\eqn\normpsibpsi{
\bra\psi(x)\bar\psi(x')\ket
\sim \frc{i\mR}{1-z'\bar z'} \ln|z-z'|^2
\indent\hbox{as}\indent |z-z'|\ra 0~,
}
as well as the condition that it vanishes in the limit of large
geodesic distance $\eta\to 1$. One can calculate the vacuum
expectation value of the energy field by point-splitting
technique. Due to resonance between the energy field
$\varepsilon(x)$ and the identity field multiplied by the mass
parameter, $m\, \Ione$, one needs one more condition to define the
energy field. This condition can be taken as
\eqn\condenergy{
\lt.\frc{d}{d m} \<\varepsilon\>\rt|_{m = 0} = 0~,
}
which gives
\eqn\energyvev{
    2R\,\bra\varepsilon\ket =
    -2\mR\(\psi(\mR)+\gamma\)-1~,
}
where $\psi(x)=d\ln\Gamma(x)/dx$ and $\gamma$ is Euler's constant.
The vacuum expectation value of the energy field is related to the
constant $k$, Eq.~\logk, appearing in the short distance expansion
of the spin-spin correlation function. This can be seen from the
spin-spin operator product expansion (OPE) in the massive Majorana
theory on the pseudosphere, which has the form
\eqn\OPEsigmasigma{
    \sigma(x)\sigma(x') =
    (2R)^{-1/4}\eta^{-1/8}\(
    \strucid(\eta,\mR)\, \Ione
+ 2R\,\eta^{1/2}\,C_\varepsilon(\eta,\mR)\,\varepsilon(x') + \ldots \),
}
where $\strucid(\eta,\mR)$ and $C_{\varepsilon}(\eta,\mR)$ are
structure functions and the dots represent contibutions coming
from descendent fields. In the Majorana theory of mass $m$ on
infinite flat space, the main property of such OPE's is that the
structure functions involved are analytic in the perturbing
parameter $m$ in some region around $m=0$ (in fact, they are
entire functions of $m$). All non-analyticities around $m=0$ of
correlation functions come from the vacuum expectation values of
the fields appearing in OPE's. In the massive Majorana theory on
the pseudosphere, analyticity of structure functions is a trivial
statement, since all correlation functions are expected to be
analytic in some region around $m=0$ (expected to be finite). This
comes from the fact that the negative curvature plays the role of
an infrared regulator.  A more useful statement is that the flat
space limit $R\to\infty$ (that is, the limit where the infrared
regulator disappears) should be well defined independently on
every term in the expansion in $m$ of the structure functions.
This gives, for the structure functions, the form
\eqn\structfct{\eqalign{
    \strucid(\eta,\mR) &= 1 + C\eta^{1/2} + O(\eta) +
    \mR\eta^{1/2} \(a\ln\eta + b + O(\eta^{1/2}\ln\eta)\)
    + O(\mR^2\eta\ln^2\eta)~,
\cr
    C_\varepsilon(\eta,\mR) &= \h + O(\eta^{1/2}\ln\eta)~.
}}
Logarithmic terms appear in the part proportional to $\mR$ in
$\strucid(\eta,\mR)$ because of the resonance between
$\varepsilon$ and $m\,{\Ione}$; from \short\ the coefficient $a$ is
equal to $-1/2$.

Clearly, the OPE \OPEsigmasigma\ shows that the constant $-r
\ln k $ involved in the short distance expansion \short\ of the
spin-spin correlation function is given by
\eqn\lnKepsilon{
-\mR\ln k = R\<\varepsilon\> + \mR b + C
}
in terms of the vacuum expectation value of the energy field and
of the constants $b$ and $C$ appearing in the structure function
$\strucid(\eta,\mR)$. These constants can be obtained from the conformal
limit $m\ra0$ \Cardy,
\eqn\sigmasigma{
\<\sigma(x)\sigma(x')\>_{m=0} = (2R)^{-1/4}
\sqrt{\eta^{1/4} + \eta^{-1/4}}~,
}
which gives $C=0$, and from the known flat space limit $k(r) \ra \mR
e^\gamma/4$ as $\mR\to\infty$ \Wu, so that $b=\ln 4$. This gives
Eq.~\logk.

It is instructive to explicitly evaluate the constant $b$
by conformal perturbation theory, thereby giving a simple
derivation of the known flat space limit. This illustrates the use
of a negative curvature as an infrared regulator. The form
\structfct~ of the structure functions imply that the constant $b$
can be calculated by perturbation theory of the two point function
$\<\sigma(x)\sigma(x')\>$ about the boundary Ising CFT, expanding
to first order in $m$:
\eqn\firstorder{\eqalign{
    & \<\sigma(x)\,\sigma(x')\> = \<\sigma(x)\,\sigma(x')\>_{m=0}
\cr& \qquad
    + \frc{2R^2m}{\pi} \int_{|\zeta|<1} \frc{d\rx''
d\ry''}{(1-\zeta\bar \zeta)^2}\;
    \Big(\<\sigma(x)\sigma(x')\varepsilon(x'')\>_{m=0}
    - (2R)^{-1}\<\sigma(x)\sigma(x')\>_{m=0}\Big)
    \cr
    & \qquad + O(m^2)~,
}}
where $\zeta = \rx''+i\ry''$ and $\bar\zeta = \rx''-i\ry''$. By simple
considerations of large geodesic distance asymptotics, the integral above
is infrared ($|\zeta|\to1$) convergent. From the result \sigmasigma\ and
from
\eqn\threepoint{\eqalign{
     \<\sigma(x)\sigma(x')\varepsilon(x'')\>_{m=0}
    =
    {(2R)^{-5/4}\over2 \sqrt{\eta^{1/4} + \eta^{-1/4}}}\Bigg\{
    &\eta^{1/4}\({|\zeta||\zeta-\eta^\h|\over |1-\eta^\h\zeta|} +
{|1-\eta^\h\zeta|\over |\zeta||\zeta-\eta^\h|}\)
\cr
+~&\eta^{-1/4}\({|\zeta-\eta^\h|\over |\zeta||1-\eta^\h\zeta|} +
{|\zeta||1-\eta^\h\zeta|\over |\zeta-\eta^\h|}\)\Bigg\}~,
}}
where $\eta$ is the projective distance \projinv\ between the points $x$
and $x'$, we find
\eqn\sigmasigmaperturb{
    \<\sigma(x)\,\sigma(x')\> = (2R)^{\frc14}
    \eta^{-\frc18} \[1+\frc12 \eta^{\frc12} + O(\eta) + \mR
    \eta^{\frc12} \(-\frc12 \ln\eta + \ln 4 +
    O\(\eta^{\frc12}\)\)
    + O(\mR^2)\],
}
which, as expected, gives $C = 0,\, b = \ln 4$.

\appendix{B}{}

In this Appendix we briefly sketch the steps for obtaining the
mode decomposition \waves. The Hamiltonian \Hcharge\ in the isometric
system of coordinates \gcoord\ reads
\eqn\isohamiltonian{
    {\bf H} = \int_{-\pi/4}^{\pi/4} \frc{d\xi_\rx}{4\pi}
(-i\psiiso,i\bpsiiso)\,
    {\cal H} \(\matrix{ \psiiso \cr \bpsiiso \cr} \)~,
}
where the Hamiltonian density is
\eqn\hdens{
    {\cal H} = \( \matrix{i\frc{d}{d\xi_\rx} & \frc{2r}{\cos(2\xi_\rx)}
    \cr \frc{2r}{\cos(2\xi_\rx)} & - i\frc{d}{d\xi_\rx} \cr} \)~.
}

The Hamiltonian is just, in the language of first quantization,
the diagonal matrix element of $\cal H$ in the state represented
by the spinor wave function $\Psi = \(\matrix{\psiiso
\cr\bpsiiso}\)$,
\eqn\innerhamiltonian{
    {\bf H} = (\Psi,{\cal H}\Psi)~,
}
where the inner product between two spinor wave functions $\Psi_1$
and $\Psi_2$ is
\eqn\inner{
    (\Psi_1,\Psi_2) = \int_{-\pi/4}^{\pi/4} \frc{d\xi_\rx}{4\pi}
    \Psi_1^\dagger(\xi_\rx) \Psi_2(\xi_\rx)~.
}

From the condition on the phases of the fermion operators
$\psiiso(\xi_\rx,\xi_\ry)^\dagger = i\psiiso(\xi_\rx,\xi_\ry)$ and
$\bpsiiso(\xi_\rx,\xi_\ry)^\dagger = -i\bpsiiso(\xi_\rx,\xi_\ry)$,
and from the condition that charge conjugation symmetry
$i\psi(\xi_\rx,\xi_\ry) \leftrightarrow \b\psi(\xi_\rx,-\xi_\ry)$
be implemented on modes by $A^\dagger_\omega \leftrightarrow
A_\omega$, the mode decomposition of the fields has the form
\eqn\wavesgen{\eqalign{
    \psiiso(\xi_\rx,\xi_\ry) &= \sum_{\omega>0} \(e^{\omega\xi_\ry}
    G_\omega(\xi_\rx) A_\omega^\dagger
    - i\,e^{-\omega\xi_\ry}\,\b{G}_\omega(\xi_\rx) A_\omega \),
\cr
    \bpsiiso(\xi_\rx,\xi_\ry) &= \sum_{\omega>0} \( e^{\omega\xi_\ry}
    \b{G}_\omega(\xi_\rx) A_\omega^\dagger
    + i\,e^{-\omega\xi_\ry}\,G_\omega(\xi_\rx) A_\omega \).
}}
The spinor wave functions
\eqn\spinor{
    S_\omega(\xi_\rx) = \(\matrix{G_\omega(\xi_\rx)\cr
    \b{G}_\omega(\xi_\rx)}\),
}
for all real values of $\omega$ and with $G_{-\omega}(\xi_\rx) = -i
\b{G}_{\omega}(\xi_\rx)$, should form a complete orthogonal set of wave
functions diagonalizing the Hamiltonian density \hdens,
\eqn\eigenspinors{
    {\cal H} S_\omega = \omega S_\omega~.
}

A set of independent spinors of the form \spinor\ diagonalizing
the Hamiltonian density with eigenvalue $\omega$ is given by
\eqn\basis{
    s_\omega^+ = \(\matrix{ g_\omega^+ \cr \b{g}_\omega^+ }\)~,
\indent
    s_\omega^- = \(\matrix{ ig_\omega^- \cr -i\b{g}_\omega^-}\)~,
}
where
\eqn\fctg{
g_\omega^\pm(\xi_{\rm x})
= e^{-i\omega\xi_\rx-i\frc{\pi}4 (\omega-1\mp2\mR)}
(1+e^{4i\xi_\rx})^{\pm \mR} F\(-\frc{\omega}2 +\frc12 \pm \mR, \pm \mR;
1\pm2\mR; 1+e^{4i\xi_\rx} \),
}
and
\eqn\negativewcond{
    \b{g}_\omega^\pm(\xi_{\rm x}) = -ig_{-\omega}^\pm(\xi_{\rm x})~.
}
The branch cut of the hypergeometric function is taken from
$-\infty$ to $1$, and the hypergeometric function is chosen to be
unity in the limit $\xi_\rx\to-\pi/4$.

The (not normalized) spinors \spinor\ can then be expressed as
\eqn\spinorbasis{
    S_\omega =
    \left\{ \eqalign{ s_\omega^+ + C_\omega s_\omega^- \ \ & (\omega>0)
\cr -s_\omega^+ +
    C_\omega
    s_\omega^- \ \ & (\omega<0) } \right.~,
}
for real constants $C_\omega$ satisfying $C_{-\omega} = C_\omega$.
For a given set of $\omega$ and given associated constants
$C_\omega$, they will form a complete orthogonal set if the inner
product $(S_\omega,S_{\omega'})$ is well defined for all $\omega$
and $\omega'$ in this set; and if the Hamiltonian is Hermitian,
$(S_\omega, {\cal H}S_{\omega'}) = ({\cal H} S_\omega,
S_{\omega'})$. These lead respectively to the conditions
\eqn\innerwelldefcomp{
    \lim_{\epsilon\to0} \epsilon \left\{ \Re e\(\b{G}_{\omega}
G_{\omega'}\)_{\xi_\rx = \pi/4-\epsilon}
    + \Re e\(\b{G}_{\omega} G_{\omega'}\)_{\xi_\rx = -\pi/4+\epsilon}
\right\}= 0~, }
and
\eqn\hermcomp{
    \lim_{\epsilon\to0} \left\{ \Im m \(\b{G}_{\omega} G_{\omega'}\)_{\xi_\rx = \pi/4-\epsilon}
    - \Im m\(\b{G}_{\omega} G_{\omega'}\)_{\xi_\rx =
    -\pi/4+\epsilon}\right\}
    = 0~.
}

In the case $\mR>\h$, condition \innerwelldefcomp\ is satisfied
only with $C_\omega = 0$ and by taking the hypergeometric function
in \fctg\ to have trivial monodromy: $\omega = \pm (1+2\mR + 2n),\
n=0,1,2,\ldots$. The function $G_\omega$ then vanishes at the
boundaries $\xi_\rx = \pm \pi/4$. With this wave decomposition,
the Fermi fields vanish as $e^{-m\dist}$ as the geodesic distance
$\dist$ to the center of the disk goes to infinity.

In the case $0<\mR<\h$, condition \innerwelldefcomp\ is always
satisfied. Condition \hermcomp\ is then satisfied for many sets
$\{\omega; C_\omega\}$. They correspond to many sets of stable
asymptotic conditions on the fields, hence to many stable regimes
of the quantum field theory with different thermodynamic
properties. In this paper we concentrate our attention on the
regimes which preserve the $SU(1,1)$ symmetry; we intend to
analyse other regimes in a future paper. The charges \charges\
must then be well defined (and the Hilbert space must form a
lowest weight module for the $su(1,1)$ algebra that they
generate), which imposes again that the hypergeometric function in
\fctg\ has trivial monodromy, but not that the function $G_\omega$
be vanishing at the boundaries $\xi_\rx = \pm \pi/4$. Hence there
are two possible sets: $\omega = \pm (1+2\mR + 2n),\
n=0,1,2,\ldots$ with $C_\omega=0$; and $\omega = \pm (1-2\mR +
2n),\ n=0,1,2,\ldots$ with $C_\omega\to\infty$. In the first set,
the Fermi fields vanish as $e^{-m\dist}$ as the geodesic distance
$\dist$ to the center of the disk goes to infinity, whereas in the
second set, they diverge as $e^{m\dist}(1+O(e^{-\dist/R}))$. These
correspond respectively to the ``fixed'' an ``free'' asymptotic
conditions on the order field~$\sigma$.

The decomposition \waves\ follows from these considerations, with
the identification $A^\dagger_\omega\mapsto
e^{i{\pi\over2}n}A^\dagger_n$ and $A_\omega\mapsto
e^{-i{\pi\over2}n}A_n$; this choice of phases insures hemiticity of the 
conserved charges \charges\ on the Hilbert space.

\appendix{C}{}

In order to derive the $SU(1,1)$ transformation properties of the
conserved charges introduced in Section~3 (Eqs.~\charges,
\Qcharge, \chargesX, \chargesY\ and \chargesZ), it is convenient
to use the fact that they can be written as integrals of local
densities quadratic in Fermi fields and their Lie derivatives.
Hence, it suffices to consider the properties of the Fermi fields
specified in Eq.~\fermitransf\ under the $SU(1,1)$ coordinate
transformation \transf. The transformation of their Lie
derivatives \Lie\ follows immediately,
\eqn\Lietransf{\eqalign{
&{\cal P}\,\psi(x) \mapsto
  (\partial f)^{-1/2}\(
  \bar a^2{\cal P}
- b^2\bar{\cal P}
- 2i\bar ab{\cal R}\)\psi(x)~,
\cr
&\bar{\cal P}\,\psi(x) \mapsto
 (\partial f)^{-1/2}\(
  a^2\bar{\cal P}
- \bar b^2{\cal P}
+ 2ia\bar b{\cal R}\)\psi(x)~,
\cr
&{\cal R}\,\psi(x) \mapsto
 (\partial f)^{-1/2}\(
  i\bar a\bar b{\cal P}
- ia b\bar {\cal P}
+ (a\bar a+b\bar b){\cal R}\)\psi(x)~,
}}
with similar expressions for $\bar\psi$ obtained by replacing
$\psi\ra\bar\psi$ and $\partial f\ra\bar\partial\bar f$.

It is then a straightforward exercise to derive the transformation
properties of the integrals of motion and hence their action on
local fields at an arbitrary point. Consider the charges
${\bf P}$, ${\bf\bar P}$, ${\bf R}$ defined in \charges. We have
\eqn\Pchargestransf{\eqalign{
&{\bf P} \mapsto \bar a^2{\bf P} + b^2{\bf\bar P} + 2\bar ab{\bf R}~,
\cr
&{\bf\bar P} \mapsto \bar b^2{\bf P} + a^2{\bf\bar P} + 2a\bar b{\bf R}~,
\cr
&{\bf R} \mapsto
  \bar a\bar b{\bf P} + a b {\bf\bar P} + (a\bar a+b\bar b){\bf R}~.
}}
Starting with their action on a local field placed at the origin,
we change coordinates with $\bar b = \bar a z_0$ and $\bar a = a =
(1-z_0\bar z_0)^{-1/2}$ in \transf\ so that the origin is taken
into $(z_0,\bar z_0)$. Eqs.~\Pchargestransf, \Otransf\ allow us to
express everything at the origin. Replacing ordinary coordinate
derivatives by the covariant ones \covder\ one is left with fields
of definite $SU(1,1)$-dimension, so that it is a simple matter to
express everything in the frame $(z,\bar z)$.

We illustrate this procedure with ${\bf P}$ acting on a local
field of $SU(1,1)$-dimension $(h,\bar h)$, and use a prime to
denote quantities in the frame $(z,\bar z)$:
\eqn\Pchargeexample{\eqalign{
[{\bf P},{\cal O}(0)]
\mapsto [{\bf P'},{\cal O}'(x)]
&=  i\,a^{2(h+\bar h+1)}\(\partial-\bar z^2\bar\partial+2(h-\bar h)\bar
z\){\cal O}(0)
\cr
&= i\,\({\cal D}'-\bar z^2\bar{\cal D}'+{2(h-\bar h)\bar z\over1-z\bar
z}\){\cal O}'(x)
\cr
&= i\,\(\partial'-\bar z^2\bar\partial'-2\bar h\bar z\){\cal O}'(x)~,
}}
in agreement with \chargeaction.

Transformation properties of remaining charges are
\eqn\Ychargestransf{\eqalign{
&{\bf Y}_1~\,\mapsto \bar a^2 {\bf Y}_1 + b^2 {\bf Y}_{-1}
+ 2\bar a b {\bf Y}_0~,
    \cr
&{\bf Y}_{-1} \mapsto \bar b^2{\bf Y}_1 + a^2{\bf Y}_{-1} + 2a\bar b{\bf
    Y}_0~, \cr
&{\bf Y}_0~\,\mapsto  \bar a\bar b {\bf Y}_1 + a b {\bf Y}_{-1} +
(a\bar a + b\bar b){\bf Y}_0~,
}}
and
\eqn\Ychargestransf{\eqalign{
{\bf Z}_0 &\mapsto {\bf Z}_0~,
\cr
{\bf Z}_1 &\mapsto
\bar a^3 \bar b {\bf Z}_2
- a b^3 {\bf Z}_{-2}
+ \bar a^2(a\bar a+3b\bar b){\bf Z}_1
- b^2(3a\bar a+b\bar b){\bf Z}_{-1}
+ \bar a b(a\bar a+b\bar b){\bf M}~,
\cr
{\bf Z}_2 &\mapsto
  \bar a^4 {\bf Z}_2
- b^4 {\bf Z}_{-2}
+ 4\bar a^3 b{\bf Z}_1
- 4\bar a b^3{\bf Z}_{-1}
+ 2\bar a^2 b^2{\bf M}~,
\cr
{\bf M}&\mapsto
3\bar a^2 \bar b^2 {\bf Z}_2
- 3a^2 b^2 {\bf Z}_{-2}
+ 6(a \bar a + b\bar b)
\[\bar a\bar b{\bf Z}_1 - a b{\bf Z}_{-1}\]
+(1+6a\bar ab\bar b) {\bf M}~,
}}
where we have introduced the combination
\eqn\Mdef{
{\bf M} = \(\mR^2-{1\over4}\){\bf Z}_0 + 3{\bf\widetilde Z}_0~.
}
Transformation properties of ${\bf Z}_{-2}$ and ${\bf Z}_{-1}$ can be
obtained from those of ${\bf Z}_2$ and ${\bf Z}_1$ by interchanging
${\bf Z}_2\leftrightarrow{\bf Z}_{-2}$, ${\bf Z}_1\leftrightarrow{\bf
Z}_{-1}$, $a\leftrightarrow\bar a$ and $b\leftrightarrow\bar b$.
Basically, ${\bf Y}_{\pm1}$ and ${\bf Y}_0$ transform in the
three-dimensional adjoint representation of $SU(1,1)$ whereas ${\bf
Z}_{\pm2}$, ${\bf Z}_{\pm1}$ and ${\bf M}$ transform in the symmetric
part of the tensor product of two three-dimensional adjoint representation.

Using the procedure described above one can check that \Qddsigmamu\
follows from
\eqn\Qddsigmamuorigin{\eqalign{
i\,\[{\bf Z}_0,\partial\bar\partial\sigma_a(0)\,\mu_b(0,0)\]
=\partial\mu_a(0)\,\bar\partial\sigma_b(0)
&+\bar\partial\mu_a(0)\,\partial\sigma_b(0)
-\partial\bar\partial\mu_a(0)\,\sigma_b(0)
\cr
&+\(\mR^2-1/4\)\mu_a(0)\,\sigma_b(0)~,
}}
whereas \Qtwosigmasigma\ follows from
\eqn\chargesactionb{\eqalign{
&[{\bf Z}_2,\sigma_a(0)\,\sigma_b(0)]
= 2\partial\mu_a(0)\,\partial\mu_b(0)
- \partial^2\mu_a(0)\,\mu_b(0) - \mu_a(0)\,\partial^2\mu_b(0)~,
\cr
&[{\bf Z}_2,\mu_a(0)\,\mu_b(0)]
= 2\partial\sigma_a(0)\,\partial\sigma_b(0)
- \partial^2\sigma_a(0)\,\sigma_b(0)
- \sigma_a(0)\,\partial^2\sigma_b(0)~,
\cr
&[{\bf Z}_1, \sigma_a(0)\,\sigma_b(0)]
= -\h\big(\partial\mu_a(0)\,\mu_b(0)
+ \mu_a(0)\,\partial\mu_b(0)\big)~,
\cr
&[{\bf Z}_1, \mu_a(0)\,\mu_b(0)]
= -\h\big(\partial\sigma_a(0)\,\sigma_b(0)
+ \sigma_a(0)\,\partial\sigma_b(0)\big)~,
\cr
&[{\bf\widetilde Z}_0, \sigma_a(0)\,\sigma_b(0)]
= [{\bf\widetilde Z}_0, \mu_a(0)\,\mu_b(0)]
= 0~.
}}

\appendix{D}{}

We present here a few terms of the asymptotic behaviour
of the functions $\chi(\eta)$ and $\varphi(\eta)$, related to the
correlation functions of spin fields by \param. Appropriate
solution to Eqs.~\phichieq {} has short distance $\eta\ra0$ behaviour
\eqn\smallphichi{\eqalign{
\varphi(\eta) &= -\ln (\mR\eta^{1/2}) -\ln(-\Omega)
+ \eta\,f_1(\Omega) + \eta^2f_2(\Omega)
+ O\(\eta^3\Omega^5\) ~,
\cr
\chi(\eta) &= \h\ln(8\mR\eta^{1/2}) + \ln(-\Omega)
+ \eta\, h_1(\Omega) + \eta^2 h_2(\Omega)
+ O\(\eta^3\Omega^5\)~,
}}
where $\eta$ is given by \projinv, $\Omega$ by \omegadef,
\eqn\smallphicoeffs{\eqalign{
f_1(\Omega) &= -{1\over 4\Omega}(1+\Omega)~,
\cr
f_2(\Omega) &= {1\over2^7\Omega^2}\Big(
4-13\Omega-14\Omega^2 + \mR^2\Omega(-1+2\Omega+8\Omega^3)
+ \mR^4\Omega(1-4\Omega+8\Omega^2-8\Omega^3)\Big),
}}
and
\eqn\smallchicoeffs{\eqalign{
h_1(\Omega) &= {1\over 4\Omega}(1+2\mR^2\Omega)~,
\cr
h_2(\Omega) &= {1\over2^7\Omega^2}\Big(
-4+13\Omega-2\Omega^2 + \mR^2\Omega(1+36\Omega+8\Omega^3)
+ \mR^4\Omega(-1-2\Omega+8\Omega^2-8\Omega^3)\Big)~.
}}
This solution has the large distance $\eta\ra1$ expansion
\eqn\largephichi{\eqalign{
\varphi(\eta) &= Ax^{1/2+\mR}F_1(x)
+A^3x^{3/2+3\mR}F_3(x)
+A^5x^{5/2+5\mR}F_5(x)
+ O\big(x^{7/2+7\mR}\big)~,
\cr
\chi(\eta) &= 4\ln\bar s
-A^2 x^{1+2\mR}H_2(x)
-A^4 x^{2+4\mR}H_4(x)
-A^6 x^{3+6\mR}H_6(x)
+O\(x^{4+8\mR}\)~,
}}
where we have introduced the notation $1-\eta=x$, $A$ is given by
\coefflarge, $\bar s$ by \magnetization,
\eqn\largephicoeffs{\eqalign{
F_1(x) &= F\(\th+\mR,\th+\mR;1+2\mR;x\)~,
\cr
F_3(x) &= {1\over12} + {(1+2\mR)^2(3+2\mR)\over64(1+\mR)^2}x
+ {(1+2\mR)^2(8+15\mR+6\mR^2)\over256(1+\mR)^2}x^2
+O\(x^3\)~,
\cr
F_5(x) &= {1\over80} + {(1+2\mR)^2(3+2\mR)\over256(1+\mR)^2}x
+ O(x^2)~,
}}
and
\eqn\largechicoeffs{\eqalign{
H_2(x) &= {1\over4} + {(1+2\mR)^2(3+2\mR)\over32(1+\mR)^2}x
+{(1+2\mR)^2(7+4\mR)\over128(1+\mR)}x^2
+ O(x^3)~,
\cr
H_4(x) &= {1\over32} + {(1+2\mR)^2(3+2\mR)\over128(1+\mR)^2}x
+ O(x^2)~,
\cr
H_6(x) &= {1\over192} + {(1+2\mR)^2(3+2\mR)\over512(1+\mR)^2}x + O(x^2)~.
}}

\appendix{E}{}

The procedure presented in Section~4 to obtain differential equations for
the spin-spin correlation functions can be easily extended to the Ising
field theory on a sphere with zero magnetic field,
\eqn\sphereaction{
{\cal A} = {\cal A}_{{\rm CFT}, c=1/2}
- {m\over2\pi}\int_{\IR^2} d^2x~ e^{\phi(x)} \,\varepsilon(x)~,
}
where ${\cal A}_{{\rm CFT}, c=1/2}$ denotes the free Majorana fermion
conformal field theory on the plane, the conformal factor reads
\eqn\spheremetric{
e^{\phi(x)} = {4R^2\over(1+z\bar z)^2}~,
}
and the energy operator is $\varepsilon(x) = i(2R)^{-1}(1+z\bar
z)(\psi\bar\psi)(x)$. Isometries of the sphere correspond to the
coordinate transformation
\eqn\sphereisom{
z\mapsto z' = f(z) = {az+\bar b\over -b z+ \bar a}~,
\indent
\bar z\mapsto \bar z' = \bar f(\bar z)
= {\bar a \bar z+ b\over -\bar b \bar z+ a}~,
}
where we can chose $a\bar a + b\bar b = 1$. Introducing the anharmonic
ratio
\eqn\sphereprojinv{
\eta  = {(z-z')(\bar z-\bar z')\over(1+z\bar z')(1+\bar z z')}
= \tan^2\(\dist(x,x')\over2R\)~,
}
related to the geodesic distance $\dist(x,x')$ between points $x$ and
$x'$, the correlation functions $G(\eta)=\bra\sigma(x)\sigma(x')\ket$,
$\tilde G(\eta)=\bra\mu(x)\mu(x')\ket$ are then found to satisfy
\eqn\sphereeq{\eqalign{
&\eta(1+\eta)\(G' G' - G'' G + \tilde G'\tilde G'-\tilde G''\tilde G\)
-(1+2\eta) \(G'G + \tilde G'\tilde G\) = 0~,
\cr
&(1+\eta)\(G' G' - G'' G - \tilde G'\tilde G'+\tilde G''\tilde G\)
- \(G'G - \tilde G'\tilde G\) = 0~,
\cr
&\eta\(\tilde G''G + \tilde G G'' - 2\tilde G' G'\)
+ \(\tilde G'G + \tilde G G'\)
 = {r^2+1/4\over(1+\eta)^2}\tilde G G~,
}}
where $r=mR$ and prime denotes derivative with respect to $\eta$ in
\sphereprojinv.

Appropriate solutions to the quadratic differential equations \sphereeq\ 
can be obtained using analysis similar to the one presented here for the 
pseudoshopere.

\listrefs

\bye